\documentclass[a4paper,12pt]{article}

\usepackage{bm}
\usepackage{graphics}
\usepackage[dvips]{graphicx}
\usepackage{latexsym,amsmath,amsfonts,amssymb}
\usepackage[latin1]{inputenc}
\usepackage[american]{babel}
\usepackage[dvips]{graphicx}
\usepackage{bbm}
\usepackage{cite}
\newcommand{\dvol}{d\text{vol}}

\newcommand{\Nugual}[1]{$\mathcal{N}= #1 $}

\newcommand{\orders}[1]{\calO \Bigl( #1 \Bigr)}

\newcommand{\di}{\mathrm{d}}

\newcommand{\ud}[2]{^{#1}_{\phantom{#1}#2}}
\newcommand{\du}[2]{_{#1}^{\phantom{#1}#2}}

\numberwithin{equation}{section}

\newcommand{\calB}{\mathcal{B}}
\newcommand{\calC}{\mathcal{C}}
\newcommand{\calF}{\mathcal{F}}
\newcommand{\calI}{\mathcal{I}}
\newcommand{\calK}{\mathcal{K}}
\newcommand{\calO}{\mathcal{O}}
\newcommand{\calM}{\mathcal{M}}
\newcommand{\calN}{\mathcal{N}}

\newcommand{\bbC}{\mathbb{C}}

\newcommand{\bbP}{\mathbb{P}}
\newcommand{\bbR}{\mathbb{R}}
\newcommand{\bbZ}{\mathbb{Z}}

\newcommand{\be}{\begin{equation}}
\newcommand{\ee}{\end{equation}}
\newcommand{\bea}{\begin{equation}\begin{aligned}}
\newcommand{\eea}{\end{aligned} \end{equation}}
\newcommand{\ml}[1]{\begin{multline} #1 \end{multline}}

\begin{document}

\begin{titlepage}

\hfill SISSA-25/2008/EP

\vspace{20pt}

\begin{center}
{\LARGE \bf Gauge/gravity duality 
and the interplay 
\vskip 15pt 
of various fractional branes}

\end{center}

\vspace{10pt}

\begin{center}
{\large  Riccardo Argurio${}^1$, Francesco Benini${}^2$, Matteo Bertolini${}^2$, \\
\vskip 10pt Cyril Closset${}^1$ and Stefano Cremonesi${}^2$}\\

\vskip 35pt

${}^1$Physique Th\'eorique et Math\'ematique and International Solvay Institutes \\
Universit\'e Libre de Bruxelles, C.P. 231, 1050 Bruxelles, Belgium \\ \vspace{0.3cm}
${}^2$SISSA/ISAS and INFN - Sezione di Trieste \\
Via Beirut 2; I 34014 Trieste, Italy
\end{center}

\vspace{20pt}

\begin{center}
\textbf{Abstract}
\end{center}

We consider different types of fractional branes on a $\bbZ_2$ orbifold of
the conifold and analyze in detail the corresponding gauge/gravity
duality.
The gauge theory possesses a rich
and varied dynamics, both in the UV and in the IR.
We find the dual supergravity solution
which contains both untwisted and twisted 3-form fluxes, related
to what are known as deformation and $\calN=2$ fractional branes
respectively.
We analyze the resulting RG flow from the supergravity perspective,
by developing an algorithm to easily extract it.
We find hints of a generalization of the familiar cascade of Seiberg dualities
due to a
non-trivial interplay between the different types of fractional branes.
We finally consider the IR behavior in several limits, where the
dominant effective dynamics is either confining, in a Coulomb phase 
or runaway, and
discuss the resolution of singularities in the dual geometric background.
\end{titlepage}


\setcounter{tocdepth}{2}

\tableofcontents

\section{Introduction}

The correspondence between gauge theories with non-trivial low-energy
dynamics and string theory backgrounds has an enormous potential.
The string theory setup is usually established drawing
uniquely on the holomorphic data of a supersymmetric gauge theory, including
a specific choice of vacuum. Then, solving the classical equations
of motion of supergravity one can in principle obtain, through
the warp factor, all the dynamical informations on the gauge theory low-energy
dynamics, that would instead usually imply precise
knowledge of the K\"ahler sector. The limitation of this procedure to supergravity
and not to full string theory corresponds in the gauge theory to taking
some large $N$ and strong 't Hooft coupling limit.

A fruitful arena where to address these issues has proven to be that of
D3-branes at Calabi-Yau (CY)
singularities. In this context, the most
celebrated example where such a program has been successfully
completed is the warped deformed conifold \cite{Klebanov:2000hb},
which describes a theory with confinement and chiral symmetry breaking. 

It is of obvious interest to apply the above program to gauge theories
with a varied low-energy behavior. D3-branes at CY singularities typically 
give rise to $\calN=1$ quiver gauge theories, which are 
supersymmetric theories characterized
by product gauge groups, matter in the bifundamental representation and a
tree level superpotential, all such data being dictated by the structure of the singularity.
Most quiver gauge theories can
have several different IR behaviors, depending on which branch
of the moduli space one is sitting on. Already in the simple
conifold theory, one has a baryonic
branch displaying confinement and a mass gap in the gauge sector,
and mesonic branches with a dynamics which is $\calN=4$ to a good
approximation. In more general quivers, other kinds of low-energy behaviors
are possible. Some quivers will actually have no vacua and display
a runaway behavior \cite{Berenstein:2005xa,Franco:2005zu,Bertolini:2005di,Intriligator:2005aw},%
\footnote{See \cite{Brini:2006ej} for some generalizations.}
but this leaves little hope of finding a regular
gravity dual. Other quivers will on the other hand contain branches
of the moduli space where the dynamics is approximately
the one on the Coulomb branch of an $\calN=2$ theory. The latter can
also be thought of as mesonic branches, albeit of complex dimension one
instead of three as in the (generic) $\calN=4$ case.

As it has been shown in \cite{Argurio:2006ny,Argurio:2007qk}, theories with both
baryonic and $\calN=2$ mesonic branches can be very interesting because they
are likely to possess, besides the supersymmetric vacua, also
metastable supersymmetry breaking vacua. The latter arise
precisely because there is a tension between the conditions
for realizing baryonic or mesonic vacua among the various nodes of the
quiver. On the gravity/string side, the metastable vacua are
associated to the presence of anti-D3 branes. They are only metastable
because they can decay through an instanton that shifts the flux
in such a way that their charge is cancelled.
Of course, a full gravity solution of such a supersymmetry breaking
vacuum would be a wonderful arena for studying quantitatively the low-energy
dynamics of such theories.

In this paper, we take a first step towards this goal. We construct the gravity
dual of the most generic gauge theory one can engineer using D3-branes at the tip
of a $\bbZ_k$ non-chiral orbifold of the conifold \cite{Uranga:1998vf}, focusing
for simplicity, but with little loss
of generality, on the case $k=2$. This singularity admits different
kinds
of fractional branes, triggering confinement or enjoying an $\calN=2$
mesonic branch and known as deformation or ${\cal N}=2$ fractional branes, respectively.
We aim at describing the backreaction of the most general D3-brane bound state. The difficulty
in doing so stems from the fact that the UV completion which corresponds
to the supergravity solution is qualitatively different in the two cases.
For deformation branes, the renormalization group (RG) flow is best described in terms of a
cascade of Seiberg dualities which increases the overall rank of the quiver
nodes towards the UV. For $\calN=2$ branes, the RG flow (which is
indeed present and also increases the ranks towards the UV \cite{Bertolini:2000dk,Polchinski:2000mx}) seems
to be better represented by some form of Higgsing \cite{Aharony:2000pp}.

It should be clear that whenever there are $\calN=2$ branes around
the IR of the gravity dual is bound to contain some singularity. This is because
open string degrees of freedom cannot completely transmute into flux. Indeed, on
the Coulomb branch we still have by definition some surviving abelian gauge group,
which cannot be described in terms of closed string degrees of freedom. This situation is similar
to the situation where one aims at describing theories with flavors. There too, flavor
degrees of freedom must be described by open strings, and hence flavor
branes must be present in the gravity dual as physical sources \cite{Karch:2002sh}.
Thus in our set up we expect to have physical sources corresponding
to $\calN=2$ fractional branes. The main difference with respect to the case of flavor branes is that
$\calN=2$ fractional branes are not infinitely extended in the Calabi-Yau.

The main results of our analysis can be summarized as follows.
We find an explicit supergravity solution describing a generic
distribution of fractional branes, both of the deformation and $\calN=2$
kind, on the orbifolded conifold, and corresponding to the UV regime of the
dual gauge theory. It describes holographically an RG flow which exactly matches
the beta functions that one can compute in the dual field theory and the expected
reduction of degrees of freedom towards the IR, which occurs through a
cascade.
We develop an algorithm to follow the RG flow of each gauge coupling
from the supergravity solution.
An interesting feature is that in this general setting there are
cascade steps that do not always have a simple interpretation in terms
of Seiberg dualities. This is due to the presence of $\calN=2$ fractional
branes, or more generally to the presence of twisted fluxes.
Nevertheless, supergravity considerations and field theory
expectations (based on the non-holomorphic beta function) exactly match.
As far as the IR regime is concerned, we perform a non-trivial consistency
check matching the field theory effective superpotential with that predicted from
the geometric background. We also provide the solution for the 3-form
fluxes and discuss the pattern of singularities resolution,
while we only set the stage for computing the exact warp factor in this case.

The paper is structured as follows.
In section \ref{sec: gauge theory} we explain our set up and introduce the
minimal geometrical data that is needed in the following.
In section \ref{sect: sugra} we present the supergravity solution
which is expected to reproduce the UV behavior of our quiver gauge theory.
We take the CY base to be the orbifold of the singular conifold, but we take into
account all the fluxes sourced by the fractional branes and compute
their backreaction on the warp factor. We then check that the result
is indeed compatible with the expected RG flow and perform a number of non-trivial gauge/gravity
duality checks. In section \ref{sec: IR} we discuss the extension of the previous solution
towards the IR, discuss the singularity structure of our solution, their resolutions, and
match the effective superpotential obtained on the two sides of the correspondence. The
appendices contain many technical data which might help in better understanding the form
of the supergravity ansatz that we solve in the main text and the geometric structure of
the orbifolded conifold CY singularity we consider.


\section{The orbifolded conifold}
\label{sec: gauge theory}

We consider in what follows an orbifolded avatar of the familiar
conifold quiver. We focus on a non-chiral $\bbZ_2$ orbifold  of the
conifold and consider the corresponding  $\calN=1$
supersymmetric quiver gauge theory obtained by  placing a bound state
of regular and fractional D3-branes at its tip. This theory has been
analyzed at great length in \cite{Argurio:2006ny}, to which we refer for
more details.

\begin{figure}[t]
\begin{center}
\includegraphics[height=3.2cm]{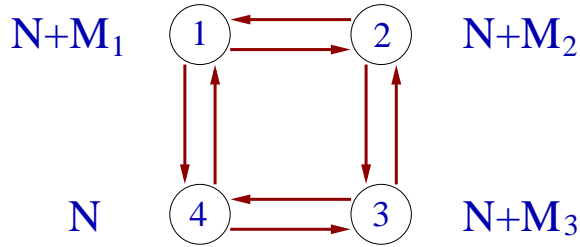}
\caption{\small The quiver diagram of the gauge theory, for the most
generic choice of ranks. Circles represent unitary gauge groups, arrows
represent bifundamental chiral superfields. For later purposes we
have parametrized the four independent ranks in terms of a common $N$. 
\label{cz2}}
\end{center}
\end{figure}

The quiver gauge theory is shown in Figure \ref{cz2}.
The gauge theory has four gauge factors and a tree level
superpotential for the bifundamental fields
\be
\label{WOrbCo}
W= \lambda \left(X_{12}X_{21}X_{14}X_{41}- X_{23}X_{32}X_{21}X_{12}
+X_{34}X_{43}X_{32}X_{23}-  X_{41}X_{14}X_{43}X_{34} \right) ~,
\ee
where $X_{ij}$ is a chiral superfield in the fundamental
representation of the $i$-th gauge group and antifundamental
representation of the  $j$-th gauge group, and traces on the gauge degrees of freedom are understood.

We are interested in the dynamics of the gauge theory with the
most generic rank assignment, as in Figure \ref{cz2}. Depending on the
values of the $M_i$'s, various kinds of IR dynamics can occur:
confinement, runaway behavior or a (locally ${\cal
N}=2$) quantum moduli space.

There is a relation between the ranks of the various gauge groups in the
quiver and the number of fractional branes wrapping the different 2-cycles
in the geometry. In turn, the fractional branes source the RR 3-form flux
which is an important ingredient in order to determine the supergravity
solution. In the following of this section we provide the link between
these three sets of data (ranks, branes wrapping cycles, fluxes).
For a more detailed discussion we refer to appendix \ref{sec: orbconifold}.

\subsection{Regular and fractional branes}

The superconformal theory ($N \not = 0\,,\,M_i=0$) can be engineered by placing
$N$ regular D3-branes at the tip of the cone. Unbalanced ranks in the quiver of Figure \ref{cz2} correspond instead
to the presence of fractional D3-branes and the corresponding breaking of conformal invariance. From the gauge
theory viewpoint, fractional branes correspond to independent anomaly free rank assignments in the quiver
(modulo the superconformal one). Hence, in the present case, we have three types of fractional branes to
play with.

In general,
fractional branes can be classified in terms of the IR dynamics they
trigger \cite{Franco:2005zu}.

A first class of fractional branes are those associated to a single node
in the quiver, or to several decoupled nodes, or else
to several contiguous nodes whose corresponding
closed loop operator appears in the tree level superpotential. This subsector of the quiver gauge theory
undergoes confinement. The dual effect in string theory is a geometric transition, which means that the
branes induce a complex structure deformation. Hence the name deformation
fractional branes. Examples of this kind  in our theory correspond to rank
assignments $(1,0,0,0)$, $(1,0,1,0)$ or $(1,1,1,0)$ and cyclic permutations.

Another class of fractional  branes are those associated to closed
loops in the quiver whose corresponding operator does not appear in
the superpotential. Such a subquiver has a mesonic moduli space
which corresponds to the Coulomb branch of an effective ${\cal N}=2$ SYM
theory. Hence the name  ${\cal N}=2$ fractional branes.
Geometrically, $\calN=2$ fractional branes are located at non-isolated
codimension four singularities in the CY three-fold. Such singularities locally look like
$\bbC \times \bbC^2/\Gamma$ (where $\Gamma = \bbZ_2$ in our case),
where the $\bbC$ complex line  corresponds to the Coulomb branch of
the effective ${\cal N}=2$ gauge theory. In the gauge theory a $U(1)^{N-1}$ gauge group survives.
In this case the branes cannot undergo a geometric transition, because there
exists no local complex deformation of such a non-isolated singularity. Hence the
supergravity dual background is expected to display some left-over singularity.
Rank assignments corresponding to this class of branes in our
quiver are for instance  $(1,1,0,0)$ and cyclic permutations.

Finally, fractional branes of any other class (which is the most
generic case, in fact) lead to ADS-like superpotential  and runaway
behavior and as such are called DSB (dynamical supersymmetry breaking) branes.
Geometrically, they are associated with geometries where the complex structure deformation is
obstructed, this tension being the geometric counterpart of the
runaway. In this case the occupied nodes have unbalanced ranks.

Obviously, combining different fractional branes of a given class, one can obtain
fractional branes of another class. Hence one can choose different fractional brane
bases to describe the gauge theory. In our present case, we will be able to
choose a basis composed only of deformation and $\calN=2$ fractional branes.
We have just seen to which rank assignments the various branes should
correspond, now we have to review which 2-cycles they are associated to.

\subsection{Geometry, cycles  and quiver ranks}\label{geom frac quiv}

There is a well established relation between quiver configurations, the primitive
topologically non-trivial shrinking 2-cycles of a given CY singularity, and the possible
existing fractional D3-branes, since the latter can be geometrically viewed as D5-branes
wrapped on such cycles. Let us review such relation for our CY singularity (see appendix \ref{sec: orbconifold}
for a full analysis).

The conifold is a non-compact CY three-fold described by
the following equation in $\bbC^4$: $z_1z_2 - z_3z_4 = 0$. We consider a $\bbZ_2$
orbifold of such singularity defined by the symmetry
\be
\label{Z2 action on z}
\Theta\,: \quad (z_1, z_2, z_3, z_4) \;\to\; (z_1, z_2, -z_3,
-z_4) ~.
\ee
The resulting orbifolded geometry is described by the following equation in $\bbC^4$
\be
(z_1 z_2)^2 - x y = 0 ~,
\ee
where $x=z_3^2$ and $y=z_4^2$. There is a singular locus in this variety which consists of two complex
lines, that we call the $p$ and $q$ lines, respectively. They meet
at the tip $\{z_1 = z_2 = x = y = 0 \}$ and correspond to the fixed point locus
of the orbifold action $\Theta$.

\begin{figure}[tn]
\begin{center}
\includegraphics[height=4.5cm]{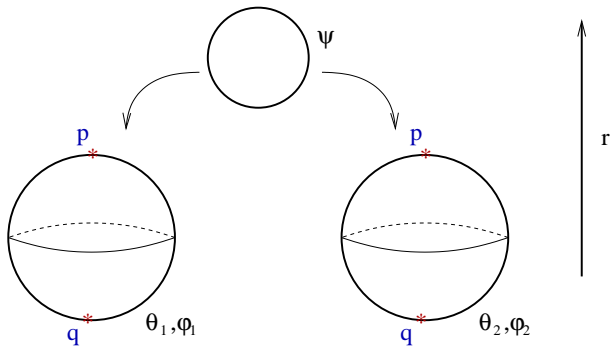}
\caption{\small The singular conifold in real angular coordinates: it
is a real cone in $r$ over $T^{1,1}$, which in turn is a $U(1)$
fibration in $\psi$ over the K\"ahler-Einstein space $\bbP^1 \times
\bbP^1$ parameterized by $\theta_i$ and $\phi_i$. The fixed point
locus of the orbifold action $\Theta$ is given by two lines $p$ and
$q$, localized at antipodal points on the two $S^2$'s. At the tip the
spheres shrink and $p$ and $q$ meet. \label{fig: t11z2} }
\end{center}
\end{figure}

One can as well describe the variety as a real manifold. The coordinates
we use are defined in appendix \ref{sec: conifold}. From this
point of view the conifold is a real cone over $T^{1,1}$, which in
turn is a $U(1)$ bundle over $S^2 \times S^2$. The orbifold action
(\ref{Z2 action on z}) reads in this case
\be
\Theta\,: \quad (\phi_1,\phi_2) \;\to\; (\phi_1 -
\pi, \phi_2 + \pi) ~.
\ee
The two complex lines are defined, in complex and real coordinates respectively, as
\bea
\label{singular lines}
p &= \{ z_1 = x = y = 0,\, \forall z_2\} = \{ \theta_1 = \theta_2 =
0,\, \forall r, \psi' \} \\ q &= \{ z_2 = x = y = 0,\, \forall z_1 \}
= \{ \theta_1 = \theta_2 = \pi,\, \forall r, \psi'' \} ~,
\eea
where $\psi'=\psi-\phi_1-\phi_2$ and $\psi''=\psi+\phi_1+\phi_2$ are
(well defined) angular coordinates along the singularity lines. In
a neighborhood of the singular lines (and outside the tip) the
geometry looks locally like the $A_1$-singularity $\bbC \times
\bbC^2/\bbZ_2$. The fixed point curve $p$
sits at the north poles of both $S^2$'s while the curve $q$ sits at
the south poles. A sketch of the conifold geometry in these real
coordinates and of the fixed points of $\Theta$ is given in Figure
\ref{fig: t11z2}.

Our CY cone has three vanishing 2-cycles. Two of these three 2-cycles arise
due to the orbifold action.
Such exceptional 2-cycles are located all along the $\bbC^2/\bbZ_2$ singular lines
$p$ and $q$, and we call them $\calC_2$ and $\calC_4$, respectively.
The third relevant 2-cycle descends from the
2-cycle of the parent conifold geometry, whose base $T^{1,1}$ is
topologically $S^2 \times S^3$.
Correspondingly, we will have a basis consisting of three fractional branes.

\begin{figure}[tn]
\begin{center}
\includegraphics[height=3.5cm]{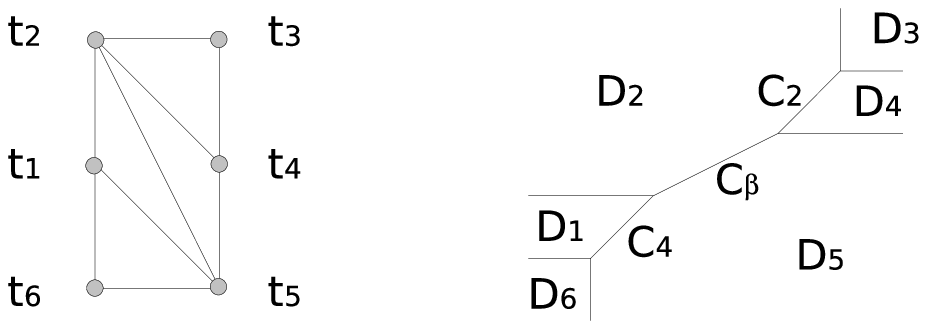}
\caption{\small The $(p,q)$-web (right) associated
to the specific triangulation (which corresponds to a specific resolution) of the toric diagram
of the orbifolded conifold (left).
\label{fig: toric orb conifold1}}
\end{center}
\end{figure}

In appendix \ref{sec: orbconifold} we construct different fractional brane
bases. However, the basis we will favor here
is the one arising most naturally when viewing our singularity as a $\bbZ_2$ projection of the conifold, which as anticipated is given
in terms of the two $\calN=2$ 2-cycles $\calC_2$ and $\calC_4$ and a deformation 2-cycle, $\calC_\beta$.
This basis of 2-cycles corresponds to a particular resolution of the singularity, which is encoded in the triangulation
of the toric diagram (and the associated $(p,q)$-web) reported in Figure \ref{fig: toric orb conifold1}.

We now mention some results derived in appendix \ref{sec: orbconifold}.
First, a
linear combination of the three cycles above, $\calC_{CF} \equiv 2\,\calC_\beta + \calC_2 + \calC_4$, has a
vanishing intersection with the exceptional 2-cycles $\calC_2$ and $\calC_4$ and it corresponds to the 2-cycle
of the double covering conifold geometry.  Hence, a brane wrapping it does not couple to closed string twisted sectors, which
are those associated to exceptional cycles, and it gives rise to
the orbifold of the configuration of a fractional brane at the singular conifold
\cite{Klebanov:2000nc}.
It thus corresponds to a quiver rank assignment $(1,0,1,0)$.
Given the obvious rank assignments $(0,1,1,0)$ and $(1,1,0,0)$ for branes
wrapped on $\calC_2$ and $\calC_4$ respectively, it follows that
the rank associated to a D5-brane wrapped on $\calC_\beta$ is $(0,-1,0,0)$. We will find it more convenient to use a 
D5-brane wrapped on $-\calC_\beta \equiv \calC_\alpha$, corresponding
to the quiver $(0,1,0,0)$.

Eventually, one needs to compute the RR 3-form fluxes sourced by each
fractional brane.
Our findings, which are derived in appendix \ref{sec: orbconifold}, are summarized in the Table below:
\be
\label{table summary frac branes}
\begin{tabular}{l|cccc}
 & $-\int_{A_2} F_3$ & $-\int_{A_4} F_3$ & $-\int_{A_{CF}} F_3$ & gauge
theory \\ \hline D5 on $\calC_2$ & $2$ & 0 & 0 & $(0,1,1,0)$ \\ D5 on
$\calC_4$ & 0 & $2$ & 0 & $(1,1,0,0)$ \\ D5 on $\calC_\alpha$ & $1$ &
$1$ & $-1$ & $(0,1,0,0)$
\end{tabular}
\ee
where fluxes are understood in units of $4\pi^2\alpha'g_s$. The 3-cycle $A_2$ corresponds to the
product of the exceptional 2-cycle $\calC_2$ transverse to the $p$-line with
the $S^1$ on $p$. Similarly, $A_4$ is the product of the
exceptional $\calC_4$ with the $S^1$ in the $q$-line. Finally, $A_{CF}$ is the
image of the compact 3-cycle of the double covering conifold under the orbifold
projection.

The table above is all we need to translate directly a quiver with
generic rank assignment to a supergravity solution with the corresponding
3-form flux.

\section{Supergravity background for the UV regime}
\label{sect: sugra}

In this section we present the supergravity solution describing the most general D3-brane system one can
consider on the orbifolded conifold. The solution is expected to be dual to the previously
discussed gauge theory with the most general rank assignment: $(N+M_1,N+M_2,N+M_3,N)$.\footnote{Our conventions for type
IIB supergravity and D-brane actions, together with the equations of motion for the bulk fields, can be found
in appendix \ref{sec: action}.}

Fractional branes are magnetic sources for the RR 3-form flux.
This typically results in some singularity of the backreacted supergravity solution.
In some cases, namely when there are only deformation branes around, the singularity
is smoothed out by the complex structure deformation the branes induce.
One gets back a singularity-free solution where branes are replaced by fluxes \cite{Klebanov:2000hb,Vafa:2000wi}.
In more general situations it is more difficult to find a regular solution. As already noticed,
in the case of ${\cal N}=2$ fractional branes this is in fact not even expected to be possible, because
there should always be some remaining open string modes corresponding to the
left over $U(1)^{N-1}$ gauge degrees of freedom on the Coulomb branch. Hence, (a remnant of) the brane sources remains in the gravity dual.

This said, in order to take the leading effect of any such kind of fractional brane into account, it is enough to make
an educated ansatz for the supergravity fields and to impose suitable boundary conditions on the system of differential equations.
Therefore, in what follows, we will only consider the type IIB bulk action $S_{IIB}$, eq.~(\ref{sugraIIB}), and implement
 the effects of each brane source by properly chosen boundary conditions.

\subsection{The UV regime: running fluxes and singularity lines}
\label{subsol}

The general solution we are looking for has constant axio-dilaton $\tau = C_0 + i e^{-\Phi} =i $, but
non-trivial RR and NSNS 3-form fluxes (which are usually organized in a complex 3-form $G_3 = F_3 + i e^{-\Phi} H_3= F_3 + i H_3$),
RR 5-form field strength $F_5$ and warp factor. The ansatz reads
\bea \label{ansatz}
ds^2_{10} &= h^{-1/2} dx^2_{3,1} + h^{1/2} (dr^2 +r^2 ds^2_{T^{1,1}}) \\
F_5 &= (1 + \ast_{10}) \, dh^{-1} \wedge \dvol_{3,1} \\
G_3 &= G_3^U + G_3^T
\eea
where the orbifold $\mathbb{Z}_2$ identification \eqref{Z2 action on z} acting on the internal coordinates is understood,
$h$ is the warp factor, while the superscripts $U$ and $T$ on the 3-form flux stand for untwisted and twisted sector
fluxes, respectively. The above ansatz is the one of a warped singular cone. Any
deformation of the singular geometry will still asymptote to this cone
for large values of the radial coordinate, and it is in this sense that we will think of the solution
as representing (at least) the UV regime of the dual gauge theory.

Recall that for the solution to be supersymmetric, the complex 3-form $G_3$ should be
$(2,1)$, primitive and imaginary-self-dual \cite{Grana:2001xn}
\be
\ast_6 \,G_3=  i \,G_3 ~,
\label{ISD}
\ee
where $*_6$ is constructed with the unwarped metric.
We will see that the warp factor depends on the radial coordinate as well as some of the angular coordinates, as
typical for solutions with ${\cal N}=2$ branes around \cite{Bertolini:2000dk}.

The equations of motion we have to solve are written in appendix \ref{sec: action}, eqs.~(\ref{eomsugra}). The warp factor
equation is given by the BI for $F_5$. The Einstein equations are then
automatically satisfied by our ansatz (\ref{ansatz}).

It is easy to check that, given all the geometrical data discussed in the previous section, and taking for simplicity
all fractional branes sitting at the tip,
the complex 3-form $G_3$ reads%
\footnote{The vielbein we use for the singular conifold can be found in \eqref{vielbein}. Appendix \ref{sec: conifold}
contains a review of the singular conifold geometry.}
\bea \label{flux N=2}
G_3 &= -\frac{\alpha'}{2} g_s\, (M_1-M_2+M_3) \,\left[ \omega_3^{CF} - 3i
\frac{dr}{r}\wedge \omega_2^{CF} \right] \\
&+ 2i\pi\alpha' g_s \,(-M_1+M_2+M_3)
\,\frac{d z_2}{z_2} \wedge \omega_2^{(p)} +2 i\pi\alpha' g_s \,(M_1+M_2-M_3) \, \frac{d z_1}{z_1} \wedge \omega_2^{(q)} \\
&=  -\frac{\alpha'}{2} g_s \,(M_1-M_2+M_3) \, \left[ \omega_3^{CF} - 3i \frac{dr}{r}\wedge \omega_2^{CF} \right] \\
& + i\pi\alpha' g_s \,(-M_1+M_2+M_3)
\,\left(3\frac{dr}{r}+i \,d\psi'\right) \wedge \omega_2^{(p)} \\
& + i\pi\alpha' g_s \,(M_1 + M_2 - M_3) \,
\left(3\frac{dr}{r}+i \,d\psi''\right) \wedge \omega_2^{(q)}~,
\eea
where  $\omega_3^{CF}$ and $\omega_2^{CF}$ are defined in appendix \ref{sec: conifold}, and $\omega_2^{(p)}$ and $\omega_2^{(q)}$ are
the two normalized exceptional 2-cocycles defined by the integrals below.

For the present purposes it suffices to recall that%
\be
\int_{\calC_{CF}} \omega_2^{CF} = 4\pi \;,\qquad
\int_{\calC_2} \omega_2^{(p)} = \int_{\calC_4} \omega_2^{(q)} = 1 \;,\qquad\text{and}\qquad \int_{A_{CF}} \omega_3^{CF} =8\pi^2~,
\ee
where $A_{CF}$ is the image under the orbifold projection of the 3-sphere on the double covering conifold.
The second equality in \eqref{flux N=2} can be easily obtained by using
eqs.~(\ref{coordC1}-\ref{coordC4}). It is then easy to check that the RR 3-form fluxes on the A-cycles are
\begin{align}
-\frac{1}{4\pi^2\alpha' g_s}\int_{A_{CF}} F_3 &= M_1 - M_2 + M_3 \\
-\frac{1}{4\pi^2\alpha'g_s}\int_{A_2} F_3 &= -M_1 + M_2 + M_3 \\
-\frac{1}{4\pi^2\alpha'g_s}\int_{A_4} F_3 &= M_1 + M_2 - M_3 ~.
\end{align}
It is important to stress at this point that the above equations are
really the input (i.e. the asymptotic conditions) in solving the equations.
They are in one-to-one correspondence with a choice of ranks in the quiver.
The real part of $G_3$, that is $F_3$, is thus essentially determined in this
way. Then the imaginary self-dual condition (\ref{ISD}) fixes also $H_3$,
the imaginary part of $G_3$. The latter is thus the output of solving the
supergravity equations. As we will see in the next subsection, this is
a non-trivial output in the sense that it will contain information
about the running of the gauge couplings. Further dynamical data on the
dual gauge theory is contained in the warp factor.

From the ansatz (\ref{ansatz}), one sees that the
warp factor should satisfy the following equation in the unwarped internal manifold
\be
\label{warph}
\ast_6 \ d \ast_6 dh \equiv \Delta h = - \ast_6 (H_3
\wedge F_3)~,
\ee
with boundary conditions dictated by the D-brane sources.
To compute $H_3 \wedge F_3$ from \eqref{flux N=2} and to solve for the warp factor $h$ in \eqref{warph}, the
first issue is whether
there are mixed terms between twisted and untwisted sectors in the expansion of such 6-form in
the cocycle basis. Let us consider a closed 2-form $\omega_2$, that represents the Poincar\'e dual
of an exceptional cycle ${\cal C}$ in any submanifold transverse to the singularity line, and $\alpha_2$ a smooth 2-form with vanishing flux on the
exceptional cycle. The 4-form $\omega_2 \wedge \alpha_2$, which would give mixed terms, vanishes at any point
but the singular one. One can then write $\omega_2 \wedge \alpha_2 = C \, \delta_4$ and compute $C$ as
\be
C = \int \omega_2 \wedge \alpha_2 = \int_{\cal C} \alpha_2 =0 ~.
\ee
This implies that
there are no mixed terms between the twisted sector and the untwisted one.
Then the 6-form $H_3 \wedge F_3$ is easily computed.
From
\eqref{flux N=2} for the 3-form fluxes, using
\begin{equation}
\begin{aligned}
\frac{dr}{r} \wedge \omega_2^{CF} \wedge \omega_3^{CF} &= -\frac{54}{r} dr \wedge \dvol_{T^{1,1}} \\
\omega_2^{(p)} \wedge \omega_2^{(p)} &=
- \frac{1}{4\pi^2} \,
\delta^{(2)}(1-\cos\theta_1,1-\cos\theta_2)
\, \sin\theta_1\, d\theta_1 \wedge d\phi_1 \wedge \sin\theta_2\,
d\theta_2 \wedge d\phi_2 \\
\omega_2^{(q)} \wedge \omega_2^{(q)} &=
- \frac{1}{4\pi^2} \,
\delta^{(2)}(1+\cos\theta_1,1+\cos\theta_2)
\, \sin\theta_1\, d\theta_1 \wedge d\phi_1 \wedge \sin\theta_2\,
d\theta_2 \wedge d\phi_2 ~,
\end{aligned}
\end{equation}
we get
\ml{
H_3 \wedge F_3 = 81 \, \alpha'^2 g_s^2 \, \frac{1}{r^6} \, \bigg\{
\frac{1}{2}(M_1-M_2 + M_3)^2 + (M_1 - M_2 - M_3)^2 \, \delta^{(2)}(1 - \cos\theta_1, 1 -
\cos\theta_2) \\
+ (M_1 + M_2 - M_3)^2 \, \delta^{(2)}(1 + \cos\theta_1, 1 +
\cos\theta_2) \bigg\} \, dr\wedge r^5\, \dvol_{T^{1,1}} ~.
}
The equation we have to solve for the warp factor is then
\ml{ \label{eq warp}
\Delta\, h = - 81\, \alpha'^2 g_s^2 \frac{1}{r^6} \bigg\{
\frac{1}{2}(M_1-M_2 + M_3)^2 + (M_1 - M_2 - M_3)^2 \, \delta^{(2)}(1-\cos\theta_1,
1-\cos\theta_2) \\ + (M_1 + M_2 - M_3)^2 \, \delta^{(2)} (1+\cos\theta_1, 1+
\cos\theta_2) \bigg\} ~.
}
Defining the angular function
\be
f(x,y) = \frac{1}{24} \sum_{(n,m)\neq
(0,0)}^\infty \frac{(2n+1)(2m+1)}{n(n+1) + m(m+1)} \, P_n(x) P_m(y) \;,
\ee
where $P_n(t)$ are Legendre polynomials, and which satisfies the differential equation
\be \label{angf}
\Delta_{ang} \, f(\cos\theta_1, \cos\theta_2) = -
\delta^{(2)} (1-\cos\theta_1, 1- \cos\theta_2) + \frac{1}{4} ~,
\ee
the solution finally reads  (see appendix \ref{laplacian conifold} for details)
\ml{ \label{warpf}
h = \frac{27 \pi \alpha'^2}{2} \frac{1}{r^4} \bigg\{ g_s N +
\frac{3g_s^2}{4\pi} \Big[ (M_1 -M_2 + M_3)^2 + (M_1-M_3)^2 + M_2^2 \Big] \Big( \log \frac{r}{r_0} + \frac{1}{4} \Big) \\
+ \frac{6g_s^2}{\pi} \Big[ (M_1-M_2 + M_3)^2 \, f(\cos\theta_1,\cos\theta_2) +
 (M_1 + M_2 - M_3)^2 \, f(-\cos\theta_1,-\cos\theta_2) \Big]
\bigg\} ~.
}

The constant terms inside the $\{ \dots \}$ in eq.~(\ref{warpf}) have been fixed in such a way that the effective D3-charge
at $r=r_0$ is $N$. This is a choice for the physical meaning one wants to give to $r_0$, as any such constant term can be
absorbed into a redefinition of $r_0$.

The above solution is not smooth, as the warp factor displays singularities at small $r$.
Moreover, as already anticipated, we expect an enhan\c con behavior to
be at work whenever there are $\calN=2$ branes in the original bound state. Similarly
to \cite{Bertolini:2000dk,Polchinski:2000mx}, the enhan\c con
radius can be defined by the
minimal surface below which the effective D3-charge changes sign.
The resolution of the singularities
has to do with the IR dynamics of the dual gauge theory. The structure of the vacua, as well as
the phases the gauge theory can enjoy,
depend crucially on the classes of fractional branes present and on the hierarchy  of the scales
$\Lambda_i$ associated to each quiver node.
Hence, the way the singularity is dealt with will change accordingly. These issues will be discussed
in detail in section \ref{sec: IR}.
Here we just want to stress that no matter the hierarchy between the dynamically generated scales
$\Lambda_i$ and the specific
fractional branes content, the above solution is a good description of the UV regime of the dual
gauge theory. In the following we will then present a number of non-trivial checks of the duality which
apply in this regime.

\subsection{Checks of the duality: beta functions and Maxwell charges}
\label{sec: checks}

In this subsection we perform some non-trivial checks of the proposed
gauge/gravity duality: we discuss the computation of gauge coupling beta functions and analyze the RG flow of our solutions
using standard techniques. 
In the following subsection we adopt a new perspective proposed in \cite{Benini:2007gx},
which is based on Page charges \cite{MarolfCB} and enables us to get stronger
predictions from supergravity.

Typically, given a supergravity background dual to a quiver gauge theory, the
knowledge of the various brane charges at any value of the radial
coordinate $r$ allows one, in principle, to extract the gauge ranks of
the dual theory at the scale $\mu$ holographically dual to
$r$. Furthermore, from the value of closed string fields,
one can learn about parameters and running couplings appearing in the
dual field theory. In theories like IIB supergravity, whose action contains
Chern-Simons terms leading to modified Bianchi identities for the
gauge invariant field strengths, different notions of charges carried
by the same fields may be introduced \cite{MarolfCB}. Following
standard techniques, we will start using the so-called Maxwell charges,
which are integrals of gauge invariant RR field strengths.

In order to specify the dictionary between the string and the gauge
sides, one needs to understand the details of the microscopic D-brane
configuration that realizes the field theory. As explained in
\cite{Polchinski:2000mx}, the idea is to match the brane charges of
the supergravity solution at some value of $r$ with the charges of a
system of fractional branes that, in the presence of the same closed
string fields as those of the supergravity solution, engineers the field
theory: in this way one reads the effective theory at the scale
$\mu$. A complication arises because the meaningful brane
configuration changes along the radial direction: when certain radial
thresholds are crossed the D3-charge of one of the effective
constituents of the system changes sign, and the system is no longer
BPS. One  has then to rearrange the charges into different BPS
constituents. The field theory counterpart is that, when one of the gauge
couplings diverges, one has to resort to a different description.

When the theory admits only deformation fractional branes, the link
between different field theory descriptions is established by Seiberg
duality. This was originally proposed and checked in the conifold
theory \cite{Klebanov:2000hb}, then applied to other singularities
\cite{Herzog:2004tr, Franco:2004jz} and even to theories with
non-compact D7-branes \cite{Benini:2007kg, Benini:2007gx}. In
\Nugual{2} solutions like the one of \cite{Bertolini:2000dk} the
procedure works also well \cite{Polchinski:2000mx}. In this latter
case, however, one expects the cascade not to be triggered by
subsequent Seiberg dualities: the correct interpretation is more along
the line of a Higgsing phenomenon \cite{Aharony:2000pp}.

The supergravity solution presented in Section \ref{subsol} is the first example of a
solution describing the backreaction of a bound state containing both deformation and
\Nugual{2} fractional branes, and hence represents an excellent
opportunity to study their interplay. One expects \Nugual{2}
fractional branes to behave as their cousins in pure \Nugual{2}
setups, and we will find good evidence that this is the case. The
novelty is that even deformation fractional branes, when probing a geometry
admitting \Nugual{2} branes, may have that kind of behavior, sometimes.

Let us first compare the gauge theory beta functions with the
supergravity prediction.  The anomalous dimensions of matter fields in
the UV are to leading order the same as in the conformal theory, $\gamma=-1/2$.
Defining
$\chi_a = 8\pi^2/g_a^2$, the four one-loop beta functions $b_a
\equiv \partial/\partial(\log\mu) \, \chi_a$ are then
\bea
\label{betagauge}
b_1 & = \frac{3}{2} \,(2M_1 - M_2)  \qquad\qquad & b_2 & = \frac{3}{2}
\,(- M_1 + 2M_2 - M_3)  \\ b_4 & = \frac{3}{2} \,(- M_1 - M_3)  & b_3
& = \frac{3}{2} \,(- M_2 + 2M_3) ~.
\eea
On the other hand,
inspection of the action of probe fractional D3-branes allows one to
find the dictionary between the gauge couplings and the integrals of
$B_2$ on the corresponding shrinking 2-cycles \cite{Kachru:1998ys, Morrison:1998cs, Klebanov:1998hh,Klebanov:1999rd}.%
\footnote{We warn the reader that such formul\ae{} are derived in \Nugual{2} orbifolds.
It is well known \cite{Strassler:2005qs} that they get corrected by
superpotential couplings in cases where the geometry is not an orbifold
of flat space. Nevertheless, the
correction is negligible in the UV of the supergravity solution.}
With the conventions laid out in appendix \ref{sec: action}, the
dictionary is easily found to be
\bea
\label{gauge-B2 dictionary}
\chi_2 + \chi_3 &=
\frac{1}{2\pi\alpha' g_s} \int_{\calC_2} B_2 \qquad\qquad & \chi_1 +
\chi_3 &= \frac{1}{2\pi\alpha' g_s} \int_{\calC_{CF}} B_2 \\ \chi_1 +
\chi_2 &= \frac{1}{2\pi\alpha' g_s} \int_{\calC_4} B_2 & \chi_1 +
\chi_2 + \chi_3 + \chi_4 &= \frac{2\pi}{g_s} ~,
\eea
with a
radius-energy relation in the UV region $r/\alpha' = \mu$, like in the
conformal case.  Recall that $\calC_{CF} = \calC_2 + \calC_4 -
2\calC_\alpha$.  

Integrating the NSNS 3-form given in eq.~(\ref{flux N=2}) one gets for
the $B_2$ field
\ml{ \label{B2}
B_2 = \frac{3}{2} \,\alpha' g_s \log \frac{r}{r_0} \Big[(M_1 -M_2
+M_3)\,\omega_2^{CF} +2\pi (-M_1+M_2+M_3)\, \omega_2^{(p)} \\
+ 2\pi (M_1+M_2 -M_3)\,\omega_2^{(q)}\Big] + \pi\alpha'
\left[ a_{CF} \,\omega_2^{CF} +4\pi ( a_2 \,\omega_2^{(p)} + a_4
\,\omega_4^{(p)}  ) \right] ~,
}
where $a_{CF}$, $a_2$, $a_4$ are integration constants. This implies
that
\bea
\label{B_2 gauge}
\frac{1}{2\pi\alpha' g_s} \int_{\calC_{CF}}
B_2  &= 3\, (M_1-M_2+M_3)\log \frac{r}{r_0} + \frac{2\pi}{g_s}\,
a_{CF}\\ \frac{1}{2\pi\alpha' g_s} \int_{\calC_2} B_2  &=
\frac{3}{2}\,(-M_1+M_2+M_3)\log \frac{r}{r_0} + \frac{2\pi}{g_s}\,
a_{2}\\ \frac{1}{2\pi\alpha' g_s} \int_{\calC_4} B_2  &=
\frac{3}{2}\,(M_1+M_2-M_3)\log \frac{r}{r_0} + \frac{2\pi}{g_s} \,
a_{4} ~.
\eea
The three integration constants $a_{CF}$, $a_2$, $a_4$ correspond to
the periods of $B_2$ at $r=r_0$, the latter having being
chosen to be the value of the holographic coordinate where the
effective D3-brane charge is $N$, see the discussion after
eq.~(\ref{angf}).  We can think of it as a UV cut-off for the dual
gauge theory, i.e. the scale where the dual UV bare Lagrangian is
defined. Then the integration constants fix, through
eqs. \eqref{gauge-B2 dictionary}, the bare couplings of the dual
non-conformal gauge theory.  It is easy to check that the logarithmic
derivatives of \eqref{B_2 gauge} give exactly the same
beta functions as the field theory computation in \eqref{betagauge}.

\

As generically happens in supergravity solutions dual to non-conformal
theories, the Maxwell D3-charge runs. It is easily computed from
eq.~\eqref{maxcharge} and \eqref{warpf} to be in our case
\be
\label{d3eff}
Q_{D3}^{}(r)= N + \frac{3 g_s}{2\pi} \left[ M_1^2 + M_2^2 + M_3^2 -
M_1M_2 - M_2M_3 \right] \, \log \frac{r}{r_0} ~.
\ee
As in \cite{Klebanov:2000hb}, the periods of $B_2$ are no
more periodic variables in the non-conformal supergravity solutions.
One should then investigate what the shift in $Q_{D3}^{}(r)$ is once
we move in the radial direction from $r$ down to $r'$, where $\Delta r = r - r'>0$ is
the minimal radius shift for which all the periods of $B_2$ on
$\calC_\alpha$, $\calC_2$, $C_4$ change by an
integer (in units of $4\pi^2\alpha'$). The shift in $Q_{D3}^{}(r)$
should then be compared against the gauge theory expectation for the
decrease of the ranks under a specific sequence of cascade steps.  What changes
after such a sequence are the ranks of the gauge groups,
all decreasing by the same integer number, the theory being otherwise
self-similar, and with the initial values of the couplings. Sometimes a cyclic permutation
of the gauge group factors is also needed, as in \cite{Klebanov:2000hb}. We will call such a sequence of cascade steps a
quasi-period.

We are now ready to check the supergravity predictions against the field theory
cascade in some simple cases with deformation fractional branes only,
where the RG flow can be followed by performing successive Seiberg
dualities.
\begin{figure}[t]
\begin{center}
{\includegraphics[height=7cm]{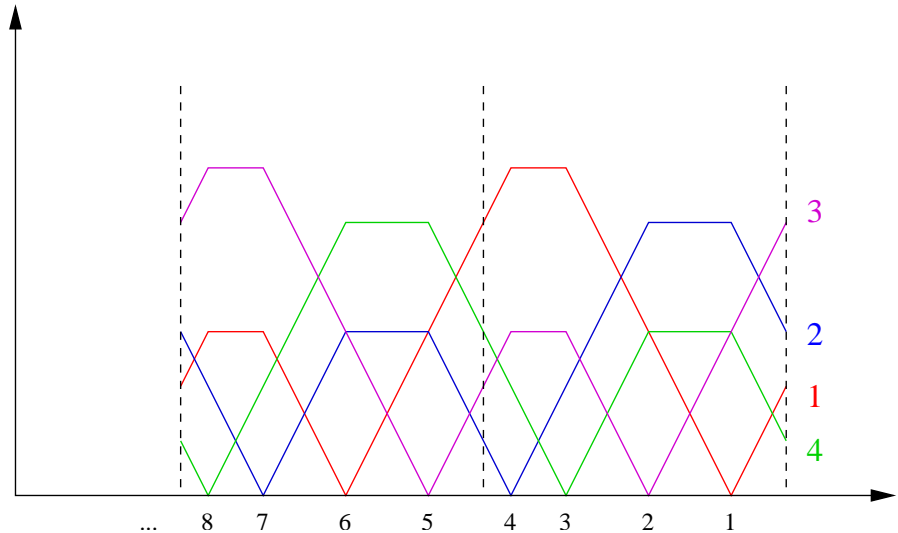}}
\caption{\small Example of the pattern of the cascade of Seiberg
dualities for ranks $(N+P,N,N+P,N)$ as derived from the field theory. 
Black numbers indicate Seiberg dualities, performed on gauge groups with diverging couplings. 
Inverse squared gauge couplings are plotted versus the logarithm of the energy scale.}
\label{P0P0cascade}
\end{center}
\end{figure}

\vskip 10pt
\noindent
{\bf 1.} $(N+P,N,N+P,N)$

This theory is the daughter of the duality cascade discussed in
\cite{Klebanov:2000hb}. There are $P$ deformation branes of type
$(1,0,1,0)$ (corresponding to D5-branes wrapped over $\calC_{CF}$). We
get for the charge and the periods
\be
\begin{split}
Q_{D3}(r) &= N + \frac{3 g_s}{4\pi} \, 4P^2  \log \frac{r}{r_0} \\
b_{\calC_\alpha} = -\frac{3 g_s}{4\pi} \,2P\, & \log
\frac{r}{r_0}+a_\alpha \;, \qquad\qquad b_{\calC_2} = a_2 \;,
\qquad\qquad b_{\calC_4} = a_4 ~,
\end{split}
\ee
where $a_{CF} = a_2 + a_4 - 2a_\alpha$ and $b_{\calC_i}$ are the periods of $B_2$ along the cycle $\calC_i$ 
in units of $4\pi^2\alpha'$. From the above equation
we see that $r' = r\, \exp[ - 4\pi/(6g_s P)]$, and under this radial
shift $Q_{D3}(r') = Q_{D3}(r) - 2P$.  This matches with the gauge
theory expectations since the theory is quasi-periodic with
a shift $N\rightarrow N-2P$, which is obtained after four
subsequent Seiberg dualities on the different gauge groups. See Figure
\ref{P0P0cascade}  for an explicit example of the RG flow computed in
field theory, for some values of the bare couplings.
Obviously, for any cyclic permutation of the above rank assignment we
have the same story.

\begin{figure}[t]
\begin{center}
{\includegraphics[height=7cm]{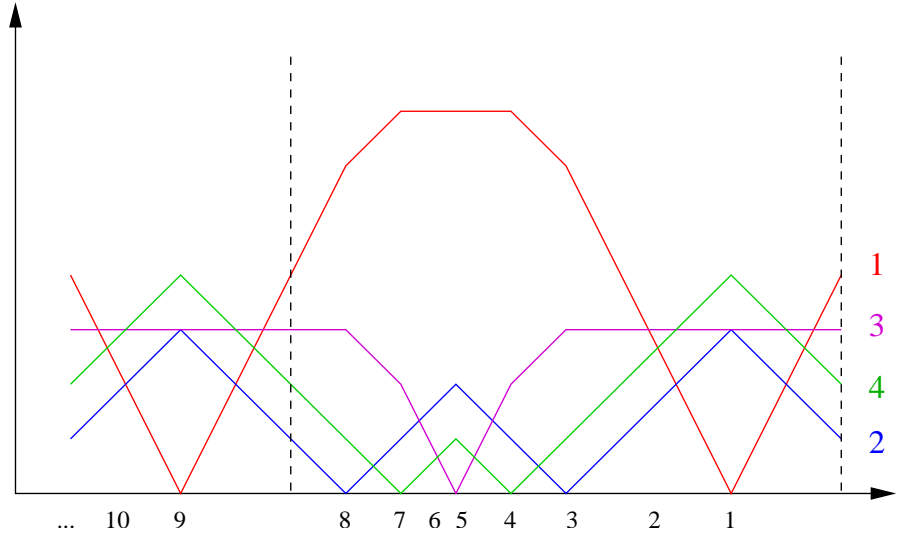}}
\caption{\small Example of the pattern of the cascade of Seiberg
dualities for ranks $(N+P,N,N,N)$ as derived from the field theory. }
\label{P000cascade}
\end{center}
\end{figure}

\vskip 10pt
\noindent
{\bf 2.} $(N+P,N,N,N)$
\be
\begin{split}
Q_{D3}(r) &= N + \frac{3 g_s}{4\pi} \, 2P^2  \log \frac{r}{r_0} \\
b_{\calC_\alpha} =  -\frac{3 g_s}{4\pi} \,P \log \frac{r}{r_0} +
a_\alpha \;, \quad b_{\calC_2} &= -\frac{3 g_s}{4\pi} \,P \log
\frac{r}{r_0} + a_2 \;, \quad b_{\calC_4} = \frac{3 g_s}{4\pi} \,P
\log \frac{r}{r_0} + a_4 ~.
\end{split}
\ee
From the above equation we see that $r' = r \, \exp[-4\pi/(3g_sP)]$ and
consequently $Q_{D3}(r')= Q_{D3}(r) - 2P$. This matches again with
gauge theory expectations. Although the quiver looks self-similar after
four Seiberg dualities, the theory is not: the gauge couplings return to their original values only after eight
Seiberg dualities, as shown in Figure \ref{P000cascade}. Hence in this
case a quasi-period needs eight dualities and the shift in the ranks
is indeed $N \rightarrow N - 2P$.
Again, similar conclusions hold for any cyclic permutations of the above rank assignment.

\vskip 10pt
\noindent
{\bf 3.} $(N+Q,N+Q,N+Q,N)$ \be
\begin{split}
Q_{D_3}(r) &= N + \frac{3 g_s}{4\pi} \, 2Q^2  \log \frac{r}{r_0} \\
b_{\calC_\alpha} = a_\alpha \;, \qquad b_{\calC_2} =  \frac{3
g_s}{4\pi} & \,Q \log \frac{r}{r_0} + a_2 \;, \qquad b_{\calC_4} =
\frac{3 g_s}{4\pi} \,Q  \log \frac{r}{r_0} + a_4 \;.
\end{split}
\ee Here, $r' = r\, \exp[-4\pi/(3g_sQ)]$ and $Q_{D3}(r')= Q_{D3}(r) -
2Q$. A quasi-period requires eight Seiberg dualities and again agrement with
field theory expectations is found.  Notice that this theory appears along the
RG flow of the theory $(N',N',N',N'+Q)$.

\subsection{Page charges and the RG flow from supergravity}

There is another way of matching our running supergravity solutions
(and more generally type IIB solutions constructed from fractional
branes at conical singularities) with cascading field theories. The method was originally
proposed in \cite{Benini:2007gx}, working on ideas in \cite{MarolfCB}.
Instead of using Maxwell charges, which are conserved and gauge invariant but not
quantized nor localized, the method is based on Page charges \cite{Page:1984qv}
which are conserved and quantized, and therefore more
suitable to be identified with gauge ranks, even though they shift
under large gauge transformations.

Let $C$ be a formal sum (polyform) of RR potentials $C = \sum C_p$,
and $F=(d + H_3\wedge)\,C$ the field strength polyform. Suppose we
have a Dp-brane, whose dual current (loosely speaking its Poincar\'e
dual) is a  $(9-p)$-form $\Omega_{9-p}$, with world-volume flux
$F_2$. Then the EOM/BI for the fluxes read
\bea
\label{EOM's polyform}
(d+H_3\wedge)\, F &=e^{\calF} \wedge \sum_p \sigma_p \; 2\kappa^2\tau_p \;
 \Omega_{9-p} \\ \Rightarrow \qquad dF^{Page} \equiv
d(e^{B_2} \wedge F) &= e^{2 \pi\alpha' F_2}
\wedge  \sum_p \sigma_p \; 2\kappa^2\tau_p \; \Omega_{9-p} ~,
\eea
where $\sigma_1=\sigma_7=1$ and $\sigma_{-1}=\sigma_3=\sigma_5=-1$.
In particular $F^{Page}$ is a closed
polyform outside the branes. Then Maxwell and Page charges are defined
as
\be
\text{Maxwell:} \quad Q_p = \frac{\sigma_p}{2\kappa^2
\tau_p} \int_{S^{8-p}} F \qquad\qquad \text{Page:} \quad Q^{Page}_p =
\frac{\sigma_p}{2\kappa^2 \tau_p} \int_{S^{8-p}} e^{B_2} \wedge
F ~.
\ee

The idea is that it is possible to read the field theory RG flow from
supergravity pointwise. At fixed radial coordinate $r$ dual to some
scale $\mu$, standard formul\ae{} allow us to compute the
gauge couplings from the dilaton and the integrals of
$B_2$. Such formul\ae{} do not give real couplings in general, but
need particular integer shifts of $B_2$, which are large gauge
transformations. Consequently, Page charges get shifted by
some integer values. Having at hand a
dictionary, they are readily mapped to the ranks of the gauge
theory at that scale.

At some specific radii, in order to keep the couplings real, 
one has to perform a further large gauge transformation,
shifting $B_2$ and therefore ending up with
different ranks. These points connect different steps of the cascade
and can usually be interpreted in the field theory as Seiberg
dualities \cite{Klebanov:2000hb} or Higgsings
\cite{Aharony:2000pp}. In particular, ranks are not continuously
varying functions but rather integer discontinuous ones. This is not
the end of the story: in general the shifts of $B_2$ are not enough to
save us from imaginary couplings, and one is forced to introduce
multiple dictionaries. We will see how everything beautifully merges.

\

Let us make the point clear using a popular example, the
Klebanov-Strassler cascade \cite{Klebanov:2000nc,
Klebanov:2000hb}. The first step is to identify a dictionary between
the field theory ranks and Page charges. An $SU(N+M) \times SU(N)$ theory is
microscopically engineered with $N$ regular and $M$ fractional
D3-branes at the tip of the conifold, thus from eq.~\eqref{EOM's polyform}
$Q_3^{Page} = N$, $Q_5^{Page} = M$. The formul\ae{} for the gauge
couplings are
\be
\chi_1 = \frac{2\pi}{g_s} \, b \qquad\qquad \chi_2 =
\frac{2\pi}{g_s} \, (1-b) ~,
\ee
where $\chi_a = 8\pi^2/g_a^2$ and
$a=1$ refers to the larger group, while $4\pi^2 \alpha' \, b =
\int_{S^2} B_2$. From the actual UV solution \cite{Klebanov:2000nc},
we have (for $B_2$ in some gauge)
\be
b = \frac{1}{4\pi^2 \alpha'}
\int_{S^2} B_2 = \frac{3g_sM}{2\pi} \log \frac{r}{r_0} \qquad\qquad
Q_3 = - \frac{1}{2\kappa^2\tau_3} \int_{T^{1,1}} F_5 = N +
\frac{3g_sM^2}{2\pi} \log \frac{r}{r_0} ~.
\ee
At any radius/energy scale $x \equiv \log r/r_0$ one should perform a
large gauge transformation and shift $b$ by some integer $\Delta b$
such that $\chi_a \geq 0$, compute the Page charges in such a gauge, and finally
use the dictionary to evaluate the ranks at that scale.

\begin{figure}[t]
\begin{center}
\includegraphics[height=5cm]{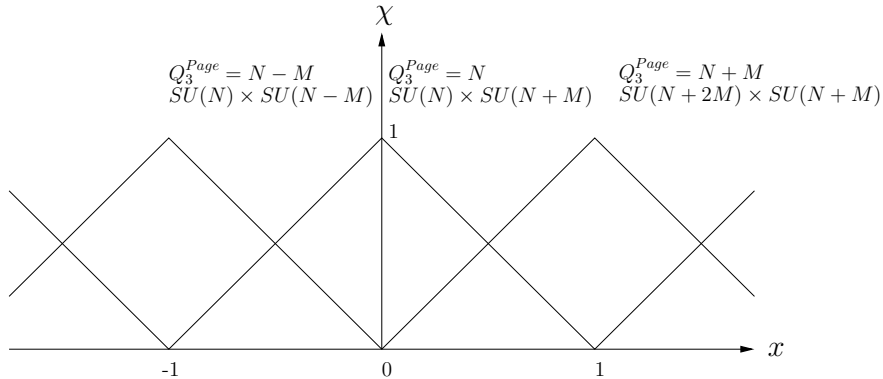}
\caption{\small Flow in the KS theory as computed with the
algorithm. $x$ is in units of $2\pi/3g_sM$ while $\chi$ in units of
$2\pi/g_s$. At integer values of $x$ a large gauge transformation is
required. At each step the Page D3-charge and the field theory is
indicated. \label{fig:KSflow}}
\end{center}
\end{figure}

It is easy to evaluate $\Delta b$ and $Q_3^{Page}$ in this example. They read
\be
\Delta b = -
\Big[ \frac{3g_sM}{2\pi} x \Big]_- \qquad\qquad Q_3^{Page} = N -
\Delta b \, M = N + \Big[ \frac{3g_sM}{2\pi} x \Big]_- \, M ~,
\ee
where the floor function $[y]_-$ is the greatest integer less than or
equal to $y$. Applying the algorithm at any $x$, we can plot the RG
flow of the gauge couplings and the ranks along it. The result
(the famous KS cascade) is depicted in Figure \ref{fig:KSflow}.
Notice that we never imposed continuity of the gauge couplings (even
though it is a well motivated physical requirement), nevertheless the
supergravity solution predicts it. Moreover it also suggests a reduction
in the gauge group ranks without explaining the corresponding
field theory mechanism. It turns out that in this case Seiberg duality
can beautifully account for it \cite{Klebanov:2000hb,Strassler:2005qs}.

\

We want to apply the same procedure to our class of solutions. In order to do
that, however, we need some more machinery. Given a basis of
2-cycles $\calC_i$ and 3-cycles $A_j$ on radial sections,
 one defines an intersection matrix
\be
C_i \cdot A_j = \calI_{ij} \qquad\qquad
i,j=1\ldots p ~,
\ee
where $p$ is the number of fractional
branes. Let $(n_I) = (\#D5_i, \#D3)$, $I=1\ldots p+1$ be the
occupation vector, that is the numbers of D5-branes wrapped on
$\calC_i$ and of D3-branes. A \emph{dictionary} $F_{(m)}$ relates this
system to the ranks $r_a$, $a=1\ldots P$ of the dual gauge theory
\be
\label{ranks from occupation}
r_a = [F_{(m)}]_{aI} \, n_I ~.
\ee
In general $P\geq p+1$, but for our non-chiral theory $P=p+1$ and
$F_{(m)}$ is invertible. In the following $i,j=1\ldots p$ while
$I,J,a,b = 1\ldots p+1$. Let $(Q_I)$ be the vector of Page charges
\be
(Q_I) = \Big( -\frac{1}{2\kappa^2\tau_5} \int_{A_j} F_3 \,,\, -
\frac{1}{2\kappa^2\tau_3} \int F_5^{Page} \Big) ~,
\ee
then the Bianchi identity eq.~\eqref{EOM's polyform} implies that $Q_j = -
\calI\ud{t}{ji} \, n_i$. Introducing the matrix $\tilde{\calI} =
\text{diag}(- \calI^t, 1)$ we can write: $Q_I = \tilde{\calI}_{IJ} \,
n_J$. It follows that (suppressing indices)
\be
\label{ranks from charges}
r = \left( F_{(m)} \, \tilde{ \calI}^{-1} \right) \, Q ~.
\ee
The formul\ae{} relating the gauge couplings to the supergravity solution
can be derived by considering the worldvolume action of probe D3- and
wrapped D5-branes \cite{Franco:2004jz}. Let $\chi_a = 8\pi^2/g_a^2$ as
before. Considering D3-branes one concludes that $\sum \chi_a =
2\pi/g_s$; then the integral of $B_2$ on some 2-cycle $\calC_j$ is
related to the gauge coupling on the probe D5-brane, which is itself
related to the sum of the $\chi$'s corresponding to the ranks increased by the 
D5, as in (\ref{gauge-B2 dictionary}). Defining the
vector
\be
(B_I) = \Big( \frac{1}{4\pi^2 \alpha'} \, \int_{\calC_i}
B_2 \,,\, 1 \Big)
\ee
one can summarize the relations by
\be
\label{chi from B}
\frac{2\pi}{g_s} \, B = F_{(m)}^t \, \chi
\qquad\qquad\Rightarrow\qquad \chi = \frac{2\pi}{g_s} \, F_{(m)}^{-1t} \, B
~.
\ee
Under large gauge transformations the integrals of $B_2$
change by integer amounts, thus the first $p$ components of the vector
$B$ undergo a particular shift $B_i \to B_i + Z_i$, for some $Z_i \in
\bbZ$. As a result the Page D3-charge is shifted by
\be
\label{delta Page D3}
\Delta Q_3^{Page} = - \frac{1}{2\kappa^2\tau_3} \int \Delta
B_2 \wedge F_3 =  Q_j \, (\calI^{-1})_{jk} \, Z_k ~,
\ee
while the inferred gauge couplings change according to eq.~\eqref{chi from B}.

\begin{figure}[t]
\begin{center}
\includegraphics[height=3.5cm]{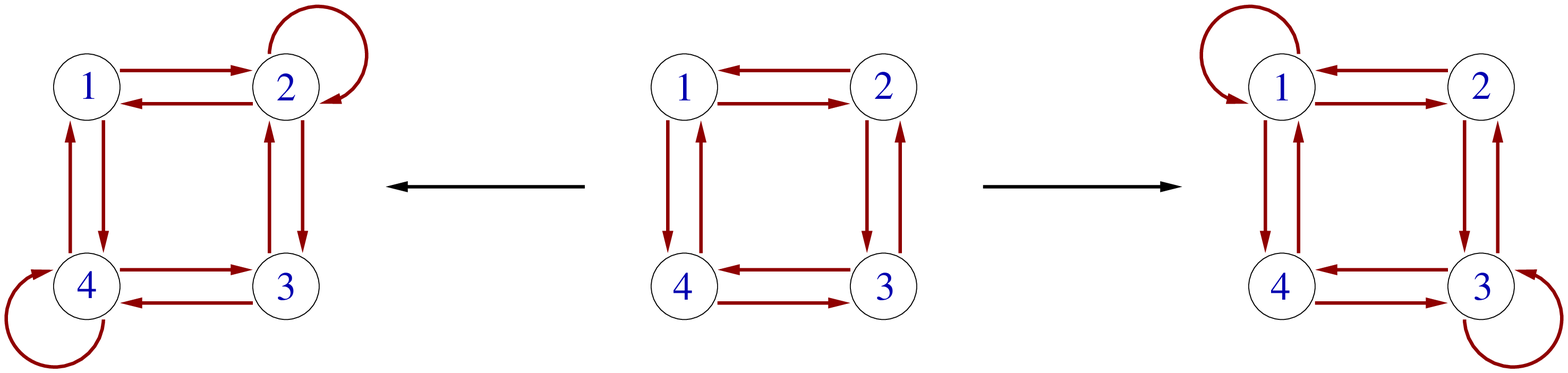}
\caption{\small Seiberg dual quivers. The central quiver is the most
extensively discussed one in the paper. The left quiver is obtained
with a Seiberg duality on node 1 or 3, while the right one on node 2
or 4. \label{fig:SDquivers}}
\end{center}
\bea \nonumber F_{(3)} = \begin{pmatrix} 0 & 1 & 0 & 1 \\ 1 & 1 & 1 &
1 \\ 0 & 1 & 1 & 1 \\ 0 & 0 & 0 & 1 \end{pmatrix} \qquad F_{(1)} =
\begin{pmatrix} 0 & 1 & 0 & 1 \\ 1 & 1 & 1 & 1 \\ 1 & 0 & 0 & 1 \\ 0 &
0 & 0 & 1 \end{pmatrix} \qquad F_{(5)} = \begin{pmatrix} 0 & 1 & 0 & 1
\\ 1 & 1 & 1 & 1 \\ 1 & 0 & 0 & 1 \\ 1 & 1 & 0 & 1 \end{pmatrix} \\
F_{(4)} = \begin{pmatrix} 1 & 0 & 1 & 1 \\ 1 & 1 & 1 & 1 \\ 1 & 0 & 0
& 1 \\ 0 & 0 & 0 & 1 \end{pmatrix} \qquad F_{(2)} = \begin{pmatrix} 1
& 0 & 1 & 1 \\ 1 & 1 & 1 & 1 \\ 0 & 1 & 1 & 1 \\ 0 & 0 & 0 & 1
\end{pmatrix} \qquad F_{(6)} = \begin{pmatrix} 0 & 1 & 1 & 1 \\ 0 & 0
& 0 & 1 \\ 1 & 0 & 1 & 1 \\ 0 & 0 & 1 & 1 \end{pmatrix} \eea
\caption{\small A set of six dictionaries for the orbifolded conifold
theory. $F_{(3)}$, $F_{(4)}$ refer to the left quiver, with adjoints
on nodes 2-4; $F_{(1)}$, $F_{(2)}$ refer to the central quiver,
without adjoints; $F_{(5)}$, $F_{(6)}$ refer to the right quiver, with
adjoints on nodes 1-3. The four columns represent the nodes activated
by a D5-brane on $\calC_2$, $\calC_4$, $\calC_\alpha$ and a D3-brane
respectively. \label{tab:dictionaries}}
\end{figure}

We now apply the algorithm to our solutions \eqref{flux N=2}, where
the integrals of $B_2$ are \eqref{B_2 gauge}, for some values of the
charges (equivalently for some $M_i$'s). Using the basis $\{\calC_2,
\calC_4, \calC_\alpha\}$ for the 2-cycles and $\{A_2, A_4, A_{CF} \}$
for the 3-cycles, the intersection matrix $\calI_{ij}$ is given by
\be
\calI_{ij} = \begin{pmatrix} -2 & 0 & 0 \\ 0 & -2 & 0 \\ -1 & -1 & 1
\end{pmatrix}
\ee
as in \eqref{table intersection numbers}, while the
dictionary $[F_{(1)}]_{aI}$ derived in section \ref{geom frac quiv}
(see Table \eqref{table summary frac branes}), referring to the
central quiver in Figure \ref{fig:SDquivers}, is reported in Figure
\ref{tab:dictionaries}. One quickly discovers that, for generic values
of the integration constants $a_i$ and of the radial coordinate $r$,
there is no gauge transformation that produces positive $\chi_a$ in
eq.~\eqref{chi from B}.

One is led to the conclusion that \emph{multiple dictionaries} are
needed. This had to be expected since performing any Seiberg duality
on the central quiver in Figure \ref{fig:SDquivers} one obtains the
lateral quivers (depending on the node chosen), which are
substantially different and cannot be described by the same
dictionary, even up to reshuffling of the nodes.

It turns out that even two dictionaries are not enough in our case. We
provide a set of six dictionaries such that, at any energy, for one
and only one dictionary there is one large gauge transformation that
gives non-negative $\chi_a$, see Figure \ref{tab:dictionaries}.

The dictionaries besides $F_{(1)}$ are obtained from it through formal
Seiberg dualities. Consider a system with occupation vector
$n=(n_1,n_2,n_3,N)$. Start with the central quiver where the ranks are
given by eq.~\eqref{ranks from occupation} using $F_{(1)}$. Then a formal
Seiberg duality on one node gives a new quiver with new ranks (and
superpotential), from which a new dictionary $F_{(m)}$ is directly
read. Actually there is an ambiguity because the number of D3-branes
$N$ could have changed in the process (but not the other charges) and
then one is free to add lines of 1's to any of the first three
columns. One can show that the physical result, that is the gauge
couplings and ranks in the correct gauge of $B_2$, is not affected. In
our case, a Seiberg duality on node 1 gives $F_{(4)}$, on node 2
$F_{(6)}$, on node 3 $F_{(3)}$, on node 4 $F_{(5)}$ and on two
opposite nodes $F_{(2)}$.

\

We can finally apply the algorithm at any radius $x \equiv \log
r/r_0$, that is:
\begin{itemize}
\item find a dictionary in the set $\{F_{(m)}\}$ and a large gauge
transformation $B_i(x) \to B_i(x) + Z_i$ such that, according to
eq.~\eqref{chi from B}, $\chi_I \geq 0 \quad \forall I$. It turns out that
there is always one and only one solution;%
\footnote{To be precise, when one of the $\chi_I$ vanishes there are
two dictionaries (with their gauges) that do the job. At these radii
there is the transition between the validity domains of two different
field theory duals.
}
\item compute the  D3-brane Page charge in this gauge, using
eq.~\eqref{delta Page D3} (D5-brane charges are invariant);
\item use the dictionary and the charges in eq.~\eqref{ranks from charges} to
evaluate the ranks at that scale in the corresponding quiver.
\end{itemize}
As a result, one can plot the gauge couplings along the flow and keep
track of the various field theory descriptions.

\begin{figure}[!np]
Figures: {\small the following figures represent the RG flow as
computed from SUGRA with the algorithm, for typical values of the
integration constants $a_2$, $a_4$, $a_\alpha$ and the initial radius
$x = \log r/r_0$. The gauge couplings are in units of $2\pi/g_s$. On
the right side we report, for each step, the dictionary used and the
ranks in the quiver;  the addition of $N$ is understood. Underlined
ranks signal an adjoint chiral superfield at the corresponding
node. The red line represents the first group, the orange the second
one, the light green the third one, the dark green the fourth one.}

\vspace*{5ex}

\begin{minipage}[T]{.6\textwidth}
\includegraphics[width=\textwidth]{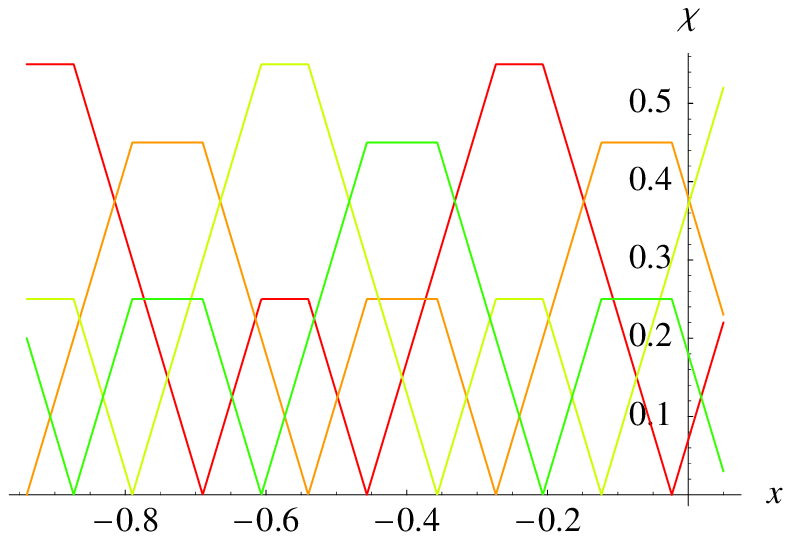}
\end{minipage}
\hfil
\begin{minipage}[b]{.37\textwidth}
\begin{tabular}{l@{(}c@{,}c@{,}c@{,}c@{)}}
\multicolumn{1}{c}{Dic.\hspace*{1ex}} & \multicolumn{4}{c}{Ranks} \\
\hline 1 & \hspace*{1pt} 1 \hspace*{1pt} & \hspace*{1pt} 0
\hspace*{1pt} & \hspace*{1pt} 1 \hspace*{1pt} & \hspace*{1pt} 0
\hspace*{1pt} \\ 4 & -1 & \underline{0} & 1 & \underline{0} \\ 2 & -1
& 0 & -1 & 0 \\ 5 & \underline{-1} & 0 & \underline{-1} & -2 \\ 1 & -1
& -2 & -1 & -2 \\ \multicolumn{2}{r}{\vdots}
\end{tabular}
\end{minipage}
\caption{\small RG flow for the $(N+1,N,N+1,N)$ theory from
SUGRA. \label{fig:Cascade1010}}

\vspace*{5ex}

\begin{minipage}[T]{.6\textwidth}
\includegraphics[width=\textwidth]{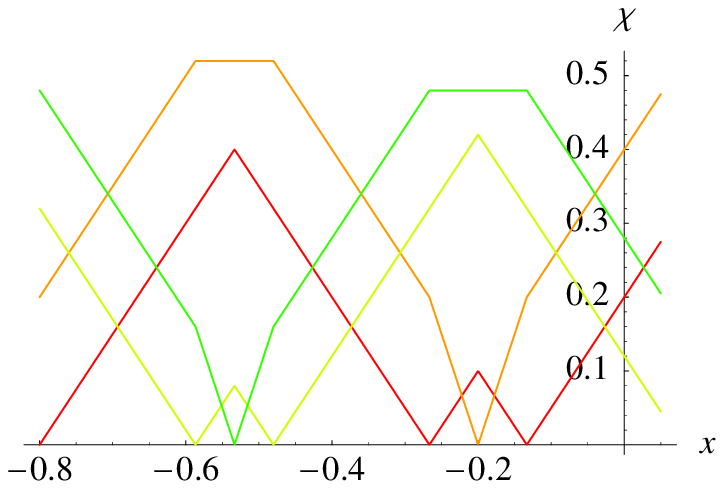}
\end{minipage}
\hfil
\begin{minipage}[b]{.37\textwidth}
\begin{tabular}{l@{(}c@{,}c@{,}c@{,}c@{)}}
\multicolumn{1}{c}{Dic.\hspace*{1ex}} & \multicolumn{4}{c}{Ranks} \\
\hline 1 & \hspace*{1pt} 1 \hspace*{1pt} & \hspace*{1pt} 1
\hspace*{1pt} & \hspace*{1pt} 0 \hspace*{1pt} & \hspace*{1pt} 0
\hspace*{1pt} \\ 4 & 0 & \underline{1} & 0 & \underline{0} \\ 5 & 0 &
\underline{-1} & 0 & \underline{0} \\ 2 & -1 & -1 & 0 & 0 \\ 6 & -1 &
\underline{-1} & -1 & \underline{0} \\ 3 & -1 & \underline{-1} & -1 &
\underline{-2} \\ 1 & -1 & -1 & -2 & -2 \\ \multicolumn{2}{r}{\vdots}
\end{tabular}
\end{minipage}
\caption{\small RG flow for the $(N+1,N+1,N,N)$ theory from
supergravity. \label{fig:Cascade1100}}
\end{figure}

\begin{figure}[!t]
\begin{minipage}[T]{.7\textwidth}
\includegraphics[width=\textwidth]{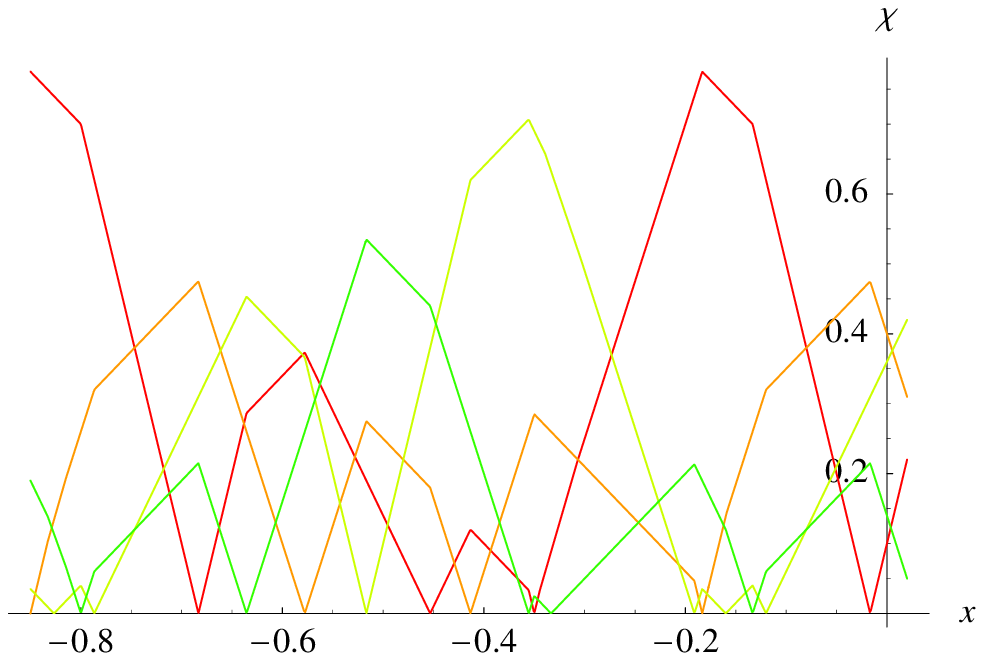}
\end{minipage}
\hfill
\begin{minipage}[b]{.27\textwidth}
\small
\begin{tabular}{l@{(}c@{,}c@{,}c@{,}c@{)}}
\multicolumn{1}{c}{Dic.\hspace*{1ex}} & \multicolumn{4}{c}{Ranks} \\
\hline 1 & \hspace*{1pt} 2 \hspace*{1pt} & \hspace*{1pt} 0
\hspace*{1pt} & \hspace*{1pt} 1 \hspace*{1pt} & \hspace*{1pt} 0
\hspace*{1pt} \\ 4 & -2 & \underline{0} & 1 & \underline{0} \\ 2 & -2
& 0 & -1 & 0 \\ 5 & \underline{-2} & 0 & \underline{-1} & -3 \\ 3 &
\underline{-2} & 0 & \underline{-2} & -3 \\ 2 & -2 & -4 & -2 & -3 \\ 6
& -2 & \underline{-4} & -5 & \underline{-3} \\ 3 & -2 & \underline{-4}
& -5 & \underline{-4} \\ 2 & -6 & -4 & -5 & -4 \\ 5 & \underline{-6} &
-4 & \underline{-5} & -7 \\ 1 & -6 & -7 & -5 & -7 \\ 3 & -8 &
\underline{-7} & -5 & \underline{-7} \\ 2 & -8 & -7 & -9 & -7 \\ 5 &
\underline{-8} & -10 & \underline{-9} & -7 \\ 1 & -8 & -10 & -9 & -10
\\ \multicolumn{2}{r}{\vdots}
\end{tabular}
\end{minipage}
\caption{\small RG flow for the $(N+2,N,N+1,N)$ theory from
supergravity. \label{fig:Cascade2010}}

\vspace*{3ex}

\begin{minipage}[T]{.6\textwidth}
\includegraphics[width=\textwidth]{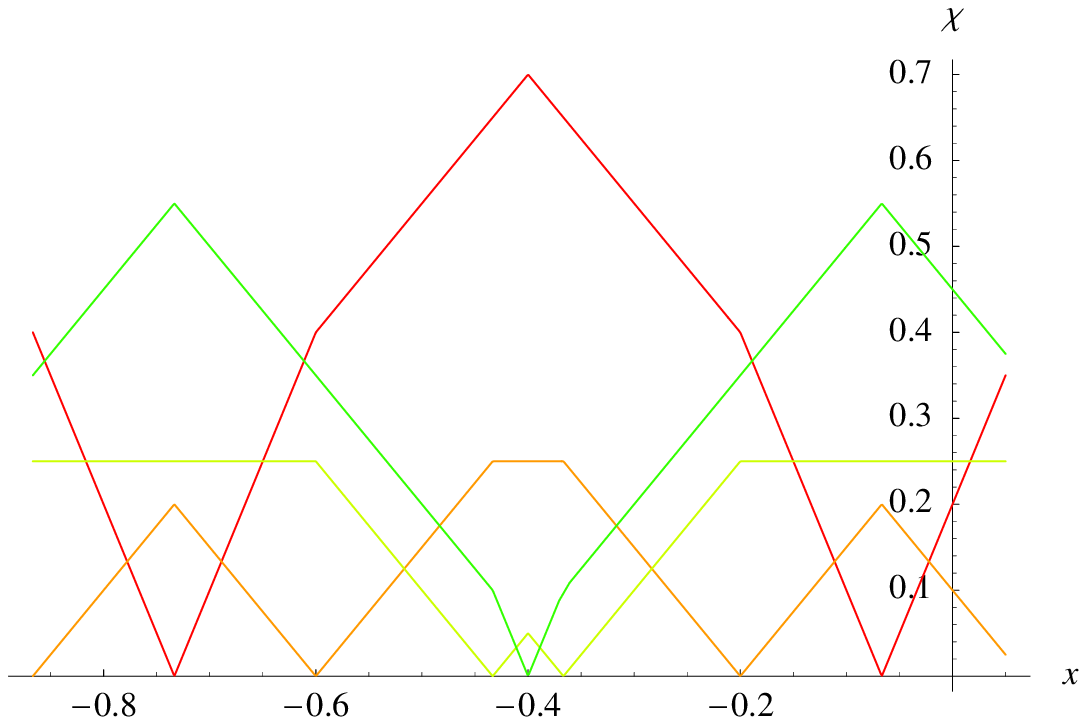}
\end{minipage}
\hfill
\begin{minipage}[b]{.37\textwidth}
\begin{tabular}{l@{(}c@{,}c@{,}c@{,}c@{)}}
\multicolumn{1}{c}{Dic.\hspace*{1ex}} & \multicolumn{4}{c}{Ranks} \\
\hline 1 & \hspace*{1pt} 1 \hspace*{1pt} & \hspace*{1pt} 0
\hspace*{1pt} & \hspace*{1pt} 0
\hspace*{1pt} & \hspace*{1pt} 0 \hspace*{1pt} \\ 4 & -1 &
\underline{0} & 0 & \underline{0} \\ 5 & -1 & \underline{-1} & 0 &
\underline{0} \\ 1 & -1 & -1 & -1 & 0 \\ 4 & \underline{-1} & -1 & \underline{-1}
& -2 \\ 6 & \underline{-1} & -1 & \underline{-2} & -2 \\ 1
& -1 & -2 & -2 & -2 \\ \multicolumn{2}{r}{\vdots}
\end{tabular}
\end{minipage}
\caption{\small RG flow for the $(N+1,N,N,N)$ theory from
supergravity. \label{fig:Cascade1000}}
\end{figure}

It is clear that the transition radii between two different
descriptions (dictionaries) occur when one of the $\chi_I$
vanishes. But in principle there is no reason why one should expect,
from the procedure above, continuous couplings at the transition
points. Surprisingly enough, it turns out that the resulting coupling
are indeed continuous. Some plots with explanation are in Figures
\ref{fig:Cascade1010}, \ref{fig:Cascade1100}, \ref{fig:Cascade2010},
\ref{fig:Cascade1000} (obtained via a mathematica code). 
In the following, we comment on interesting examples.

\vskip 10pt
\noindent
{\bf 1.} $(N+P,N,N+P,N)$

The RG flow, as computed from supergravity with the algorithm above,
is plotted in Figure \ref{fig:Cascade1010} (for $P=1$ and some typical
choice of the integration constants $a_2$, $a_4$, $a_\alpha$ and the
starting radius $x = \log r/r_0$). It precisely matches with the field theory
expectations, with respect to both gauge couplings and ranks at any
step. All transition points can be interpreted by means of a single Seiberg duality,
as the prototypical example in \cite{Klebanov:2000hb}. Notice that the
integral of $B_2$ on $\calC_2$ and $\calC_4$ is constant and
generically not integer.

\vskip 10pt
\noindent
{\bf 2.} $(N+P,N+P,N,N)$

The supergravity RG flow is shown in Figure \ref{fig:Cascade1100} (for
$P=1$ and typical integration constants). This theory is realized with
\Nugual{2} fractional branes only, and one expects a behavior quite
similar to the \Nugual{2} setup of \cite{Bertolini:2000dk}. The
algorithm confirms that there are steps of the cascade where the node
with divergent coupling has an adjoint chiral field and \Nugual{2}
superpotential. In the example of Figure \ref{fig:Cascade1100}, after
a Seiberg duality on node 1, one is left with the left hand side quiver of
Figure \ref{fig:SDquivers}, and superpotential 
\be 
W = -X_{12}X_{21}X_{14}X_{41} + M_{22} ( X_{21}X_{12} - X_{23}X_{32} ) +
X_{32}X_{23}X_{34}X_{43} - M_{44} ( X_{43}X_{34} - X_{41}X_{14} ) ~.
\ee 
The next node with diverging coupling is node 2. Notice that if
one neglects the gauge dynamics on the other nodes and possible
subtleties related to a non-trivial K\"ahler potential and anomalous
dimensions of node 2, the theory is effectively \Nugual{2} massless
SQCD with $N+P$ colors and $2N$ flavors. One is tempted to think
that this piece of the RG flow can be interpreted as in the \Nugual{2}
theory of \cite{Bertolini:2000dk,Aharony:2000pp} 
(see also \cite{Petrini:2001fk}).

It is beyond the scope of this paper to fully understand the field
theory dynamics. We just want to observe that on the gravity
side this step in the cascade, possibly understandable as Higgsings,
precisely occurs when
\be
\frac{1}{4\pi^2\alpha'} \int_{\calC_2} B_2 \in \bbZ
\qquad\qquad\text{or}\qquad\qquad \frac{1}{4\pi^2\alpha'}
\int_{\calC_4} B_2 \in \bbZ
\ee (in this case only $\calC_4$). Since $\calC_2$ and $\calC_4$ are shrunk
2-cycles along the \Nugual{2} singularity lines, at these radii (called generalized enhan\c
cons in \cite{Aharony:2000pp}) there are extra massless fields and
tensionless objects in supergravity.

\vskip 10pt
\noindent
{\bf 3.} $(N+P,N,N+Q,N)$

The supergravity RG flow for the case $(N+2,N,N+1,N)$ is shown in
Figure \ref{fig:Cascade2010}. This theory is realized with deformation
fractional branes only. Nevertheless, the fact that the geometry admits
\Nugual{2} fractional branes causes that, at some steps, there is a
reduction of rank in a node with adjoint; as before, this cannot be
interpreted as a Seiberg duality and some other mechanism, such as
Higgsing, should be invoked. Shells where such transitions occur are
precisely at radii where one of the periods of $B_2$ on $\calC_2$ or
$\calC_4$ vanishes.

This rather intriguing fact can be understood by noticing that
in some intermediate steps, i.e. when there are nodes with adjoints,
the  relevant dictionary  forces us to reinterpret the configuration
as if it were composed of deformation fractional branes together with a
number of ${\cal N}=2$ fractional branes.

For generic $P$ and $Q$ things can be analysed in a similar way.
Notice that for $P$ and $Q$ large and coprime, the flow
becomes quickly very complicated.

\vskip 10pt
\noindent
{\bf 4.} $(N+P,N,N,N)$

The supergravity RG flow for the case $(N+1,N,N,N)$ is shown in Figure
\ref{fig:Cascade1000}. As in the previous examples, when one of the
periods of $B_2$ on $\calC_2$ or $\calC_4$ vanishes supergravity predicts
some transition that cannot be interpreted as a Seiberg duality in the
FT. This flow is anyway peculiar because performing a Seiberg duality
on a conformal node it is possible to provide a dual FT interpretation
of the RG flow using only Seiberg dualities, as was done in the previous
subsection. However, supergravity seems to predict a different pattern
of dualities which nevertheless leads to the same evolution of the gauge
couplings.

\

Let us summarize what we found. There exists a
well-defined algorithm that, given a minimal set of
dictionaries, allows one to derive the field theory RG flow from a supergravity
solution. For toric singularities, as the one we are describing, the dictionaries can 
be derived using standard techniques (see for 
instance \cite{Butti:2006hc}) and,
given the first, the other ones follow applying formal Seiberg dualities.
It is not clear to us how to determine the minimal number of dictionaries, and we
have obtained them by hand. Moreover, it would be interesting to understand how to extend the 
algorithm to supergravity solutions dual to chiral gauge theories, as those in \cite{Herzog:2004tr}.

Our geometry admits both deformation and \Nugual{2} fractional
branes. We saw examples of cascades from deformation branes that can
be interpreted in term of Seiberg dualities only, examples with \Nugual{2} branes that are very
close to pure \Nugual{2} theories and whose interpretation should be
similar to the Higgsing proposed of \cite{Aharony:2000pp}, but also examples
which one would say are realized with deformation branes only that require something
like a Higgsing, at some steps. We expect to explore this field theory
interpretation in a forthcoming paper \cite{wip}.


\section{The IR regime of the theory}
\label{sec: IR}

As already noticed, the solution presented in the previous section is singular. 
In this section we discuss how to extend it towards the IR (i.e. at small radii on the gravity side).
It is not difficult to see that the warp factor (\ref{warpf}) becomes singular at short distances, 
so that the metric has a repulson type singularity.

This is of course expected, since our solution is similar to the ones of \cite{Klebanov:2000nc} and \cite{Bertolini:2000dk}: we are considering the backreaction of the branes in the supergravity limit, but supergravity cannot be the full story near the branes themselves, where the stringy dynamics should be dominant.
Resolving the singularity then amounts to a clever guess of what these stringy effects would lead to. Deformation fractional branes and $\calN=2$ fractional branes are very different in that respect. 

In the case of deformation branes at conifold points, the singularity can 
be smoothed out in supergravity by considering the warping of the deformed conifold
instead of the singular conifold. This is what has been done in 
\cite{Klebanov:2000hb}, and the procedure introduces a dimensionful 
parameter $\epsilon$, related to the dynamical scale of a confining
gauge group. 

In the case of $\calN=2$ fractional branes, one does not expect that the repulson singularity can be smoothed in a similar way. 
Indeed, the $U(1)^{N}$ abelian degrees of freedom on the Coulomb branch can only appear through the presence of left-over open string modes in the
gravity dual. This means that physical branes are still 
present, although they are expected to form a ring that effectively cloaks the singularity \cite{Bertolini:2000dk}. 
This is the enhan\c{c}on mechanism first discussed in \cite{Johnson:1999qt}.
The enhan\c{c}on radius (where probe fractional branes become tensionless)
then provides a dimensionful parameter,
which basically corresponds to the dynamically
generated scale of the $\calN=2$ gauge theory.

Note that in addition to the repulson singularity, the presence of twisted flux makes the warp factor
singular all along the Coulomb branch, which coincides with the line
of orbifold singularities. It then signals that one should include new massless modes 
in the low energy effective theory also at large values of $r$. 
This is what happens in our $\calN=1$ orbifolded conifold setup as well.
Still, the supergravity solution can already give us some important insight into the dynamics, particularly 
about the RG flow trajectory of the gauge theory dual, as we saw in the previous section.

We now turn to the IR effective theory at the bottom of the cascade.
In our solutions, it is clear that the IR behaviour can be quite
different depending on which dynamics dominates, i.e. which nodes in
the low-energy quiver have the largest dynamical scale. As was argued
in the previous section, the RG flow will, in a way or another, reduce
the ranks  of the gauge groups by a common additive factor. In other
words, the effective number of regular branes will diminish as we go
inwards to the IR, and we  assume that we eventually reach a point
where the quiver has only three nodes.

In the following,  we will first analyze the low-energy dynamics from
the gauge theory point of view. We perform the analysis in two
different regimes: either the $\calN=2$ effective dynamics is the most
important effect, or else the $\calN=1$ confining behavior dominates.
As a consistency check of the candidate gravity dual,  we reproduce
the effective superpotential from the holomorphic data of the geometry
in that latter limit.

We eventually consider the equations determining the warp factor. The
latter is related to data encoded in the full K\"ahler potential
of the  gauge theory. Hence, computing the warp factor would be the
main challenge in order to gather new dynamical information on the
low-energy theory. To do that, the two limits in which the
dynamics is predominantly confining or $\calN=2$ are quite
different. In the latter case, we will argue that  the enhan\c{c}on is
so large that a possible local deformation of the geometry would be
irrelevant, and so the UV solution presented in the  previous section
is basically the correct gravity dual up to the enhan\c{c}on radius.
When instead the confining dynamics is the strongest,  one
expects to have a gravity dual consisting of the orbifold of the
deformed conifold, with singularities along the orbifold fixed
line. We must anticipate that we will stop short of actually computing
the warp factor in that case.

\subsection{Gauge theory IR dynamics}

\begin{figure}
\begin{center}
{\includegraphics[height=1.6cm] {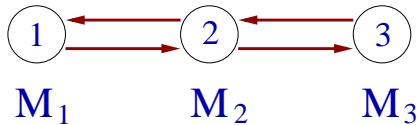}}
\caption{\small The 3-node quiver that corresponds to the IR bottom of
the cascade.}
\label{3node}
\end{center}
\end{figure}

In this subsection we perform the gauge theory analysis for the
low-energy behavior of a generic 3-node quiver, see Figure \ref{3node}.
It will often prove useful to actually think of moduli spaces in terms
of mobile (fractional) branes, so we will freely make reference to
this interpretation even in the course of the purely gauge theoretic
analysis.

Let us call $\Lambda_i$ the dynamically generated scale of the $i$-th
node of the quiver, with $i=1,2,3$. We consider two qualitatively
different regimes.

First we analyze the regime $\Lambda_2 \gg \Lambda_{1,3}$, where the
dominant quantum effects come from the second node. As we will see,
for $M_2<M_1+M_3$, there is no deformation of the (mesonic) moduli
space, which itself corresponds to having a stack of $\calN=2$
fractional branes on their Coulomb branch. For $M_2>M_1+M_3$, we find
a runaway behavior on the Coulomb branch. This is interpreted in the
gravity dual as a fully regular deformation of the geometry in the
presence of $\calN=2$ fractional branes. Indeed, in this case the
exceptional cycle the branes wrap is blown-up and minimizes its volume
at infinity: the $\calN=2$ branes are pushed away.

Secondly, the regime $\Lambda_2 \ll \Lambda_{1,3}$ is analyzed (a
similar analysis was performed in the appendix of
\cite{Argurio:2006ny}). One finds gaugino condensation for both nodes
one and three, with $S_1=S_3$. On the dual geometric side, the
deformation branes trigger a geometric transition that still preserves
an orbifold singularity line in the resulting deformed geometry. The
singularity line can accommodate some left over ${\cal N}=2$ branes
which explore their moduli space.

\subsubsection{Regime $\Lambda_2 \gg \Lambda_{1,3}$}

In this regime, the only gauge dynamics we take into account is the
one of the second node.  The quiver configuration is $(M_1,M_2,M_3,0)$
with $3 M_2 > M_1+ M_3$, so that node 2 has a strongly coupled IR
dynamics  and it makes sense to neglect the scales of the other nodes
as a first approximation.

The tree level superpotential is  
\be
\label{WtreeMPQ}
W_{tree}= \lambda X_{12}X_{23}X_{32}X_{21}   \ee  and the quantum
corrected one is  \be  W= W_{tree} - (M_1-M_2+M_3)
\left(\frac{\det{\calK}}{\Lambda_2^{3M_2-M_1-M_3}}
\right)^{\frac{1}{M_1-M_2+M_3}}~.   
\ee   
If  $M_2>M_1+M_3$, this is the familiar Affleck-Dine-Seiberg (ADS) superpotential \cite{ADS}, while if
$M_2<M_1+M_3$, it is the effective superpotential  for the free
Seiberg dual mesons and vanishing dual quark VEVs.\footnote{ In
principle, we should worry about additional baryonic directions in the
effective dynamics. Their fully quantum analysis is beyond the scope
of the present paper, however both the classical gauge theory analysis
of the higgsing patterns and their  interpretation in terms of brane
motions hint that the statements concerning the mesonic VEVs should
not be modified.}

The meson matrix for the second node is 
\be
\label{MesmatK}
\calK \equiv \bordermatrix{&    &           \cr &X_{12}X_{21}&
X_{12}X_{23} \cr &X_{32}X_{21}    & X_{32}X_{23} \cr } \equiv
\bordermatrix{&    & \cr &K_{11}& K_{13}    \cr &K_{31}    & K_{33}
\cr }~.   
\ee  
Let us denote 
\be
\label{defXi}
S_2 \equiv \left(\frac{\det{\calK}}{\Lambda_2^{3M_2-M_1-M_3}}
\right)^{\frac{1}{M_1-M_2+M_3}}~.  
\ee 
We want to determine the moduli
space of such a theory. Considering the effective superpotential in
terms of the mesons, one has the following F-flatness conditions  
\be
S_2(\calK^{-1})_{11}= 0  =  \ S_2(\calK^{-1})_{33} \nonumber 
\ee 
\be
\lambda K_{31}- S_2(\calK^{-1})_{31} = 0 = \ \lambda K_{13}-
S_2(\calK^{-1})_{13}    \ee This implies \be M_1 S_2 = M_3 S_2~.  
\ee
We must then have\footnote{Unless $M_1=M_3$, where we have another
possible solution: $K_{11}=K_{33}=0$ and  $S_2= \Lambda_2^3
(\lambda\Lambda_2)^{\frac{M_1}{M_2-M_1}}$, presumably related to a
non-Coulomb branch.}  that $S_2=0$, which implies that $K_{13}$ and
$K_{31}$ must vanish, and $\det\calK= \det{K_{11}} \det{K_{33}}$. When
$M_2<M_1+M_3$ the constraint $S_2=0$ means that $\det{\calK} =
0$. Using the gauge freedom of the first and third nodes, the general
solution consists of $\calK$ diagonal with $M_2$ non-vanishing
eigenvalues. There are as many distinct such  solutions as there are
possibilities of choosing $M_2$ out of the $M_1+M_3$ $\calN=2$
subquiver configurations $(1,1,0,0)$ or $(0,1,1,0)$.

When we have instead $M_2>M_1+M_3$, there is an ADS superpotential,
and the constraint on the mesons become 
\be \det{\calK} =
\det{K_{11}} \det{K_{33}} \rightarrow \infty  ~.  
\ee 
This corresponds to a runaway behavior of the $\calN=2$ brane configuration (the same phenomenon was observed in \cite{Berenstein:2003fx, Imeroni:2003cw}). Indeed,
after all the  $\calN=2$ configurations have been accounted for (by
moving on the Coulomb branch), there remains the configuration
$(0,M_2-M_1-M_3,0,0)$, that confines, and we know that this should
correspond to the following deformation of the geometry seen by
D3-branes  
\be 
(z_1z_2-S_2)z_1z_2 = xy~.   
\ee  
This space only has a singularity at the origin, so that the Coulomb branch (which
corresponds to a singularity line in  the orbifolded conifold) is lifted, the
supersymmetric vacua being preserved only at infinity. Geometrically
what happens is that the $\calN=2$ branes become non-BPS, as they wrap
a blown-up  cycle, and they can only minimize their tension by moving
off to infinity.

\subsubsection{Regime $\Lambda_2 \ll \Lambda_{1,3}$}\label{regimeL2llL13}

In the regime $\Lambda_{1,3}\gg \Lambda_2$, gaugino condensation at
the first and third nodes is the dominant effect in the IR. This
corresponds to a complex structure deformation of the geometry,
induced by the deformation fractional branes. We again consider the
quiver configuration $(M_1,M_2,M_3,0)$ with tree level superpotential
(\ref{WtreeMPQ}).

Let us restrict to the case where $M_2 < M_1,M_3$.\footnote{ If $M_2 >
M_1,M_3$, two Seiberg dualities on nodes one and three bring us back
to the case analyzed previously because we can assume that the dual
scales are such that  $\tilde \Lambda_{1,3}\ll \Lambda_2$. If
$M_1>M_2>M_3$, it is possible to show that the system has a runaway
behavior.}  The first and the third gauge groups develop an ADS
superpotential at the quantum level, while the second gauge group can
be considered classical. In term of the mesons ${\cal M}=X_{21}X_{12}$
and ${\cal N}=X_{23}X_{32}$ of the first and third nodes respectively
(which are both $M_2\times M_2$ matrices in the adjoint plus singlet
of the second node), the full effective superpotential reads 
\be 
W =
\lambda {\cal M} {\cal N} + (M_1-M_2)\left (
\frac{\Lambda_1^{3M_1-M_2}}{\det {\cal M}} \right)^\frac{1}{M_1-M_2}
+(M_3-M_2)\left ( \frac{\Lambda_3^{3M_3-M_2}}{\det {\cal N}}
\right)^\frac{1}{M_3-M_2}~.  
\ee 
Instead of solving for the extrema of
the above superpotential, we find it useful to first integrate in the
glueball superfields for the two confining gauge groups. We are also
motivated in doing this by the approach which uses the Gukov-Vafa-Witten (GVW) 
\cite{Gukov:1999ya} 
superpotential to make the link between the gauge theory and the
geometrical quantities, and which will be pursued in section
\ref{GVWW}.  We thus obtain 
\be 
\label{superWMN13} 
W = \lambda {\cal
M} {\cal N} + (M_1-M_2)S_1 - S_1 \log \frac{S_1^{M_1-M_2} \det
\calM}{\Lambda_1^{3M_1-M_2}} + (M_3-M_2)S_3 - S_3 \log
\frac{S_3^{M_3-M_2} \det \calN}{\Lambda_3^{3M_3-M_2}}~,  
\ee 
which is a Taylor-Veneziano-Yankielowicz (TVY) \cite{Taylor:1982bp} kind of superpotential. Of
course, extremizing with respect to $S_1$ and $S_3$ will lead us back
to the previous ADS-like superpotential. However let us extremize with
respect to all fields together 
\be
\label{MeqNmin1} 
\lambda \calN = S_1
\calM^{-1}, \qquad \qquad \lambda \calM = S_3 \calN^{-1}, \ee  \be
\log \frac{S_1^{M_1-M_2} \det \calM}{\Lambda_1^{3M_1-M_2}} = 0, \qquad
\log \frac{S_3^{M_3-M_2} \det \calN}{\Lambda_3^{3M_3-M_2}} = 0~.  
\ee
The above equations imply that $\calM$ is proportional to the inverse
of $\calN$, and that   
\be S_1=S_3\equiv  S = \left( \lambda^{M_2}
\Lambda_1^{3M_1-M_2}\Lambda_3^{3M_3-M_2}\right)^\frac{1}{M_1-M_2+M_3}~.
\ee 
This of course implies that also $\det \calM$ is fixed, while the
moduli space is spanned by the values of $\calM$ subject to this
constraint.  Once the effective $\calN=2$ dynamics of the $SU(M_2)$
gauge group is taken into account, the moduli space reduces to the
$M_2-1$ directions in the Cartan subalgebra.

Let us also consider two limiting cases. If $M_1=M_2=M_3 \equiv M$,
one can check that the mesonic and the baryonic branches decouple. On
the mesonic branch, the superpotential (\ref{superWMN13}) is correct
and the solution to its extremization is 
\be 
\det \calM =
\Lambda_1^{2M} , \qquad \det \calN = \Lambda_3^{2M} , \qquad S_1=S_3=
\left(\lambda^M \Lambda_1^{2M}\Lambda_3^{2M}\right)^\frac{1}{M}~.  
\ee
The dynamics is essentially the same as before. Note that the $S_i$
act effectively as Lagrange multipliers, and their being non zero is a
signal of the decoupling of the mesonic from the baryonic branch. This
was the case of most interest in \cite{Argurio:2006ny}.

The other limiting case is $M_2=0$. Here there are no mesons $\calM$
and $\calN$, and hence no coupling between nodes one and three.  We
just have a sum of two Veneziano-Yankielowicz superpotentials
\cite{Veneziano:1982ah} for two decoupled SYM theories.  Consistently,
we obtain upon extremization  \be S_1=\left(\Lambda_1^{3
M_1}\right)^\frac{1}{M_1}, \qquad  S_3=\left(\Lambda_3^{3
M_3}\right)^\frac{1}{M_3} ~.  \ee  In this case, the two VEVs $S_i$
are independent. It corresponds to a generic deformation of the
geometry, as reviewed in  appendix \ref{sec: 3-cycles}.

\subsection{The Gukov-Vafa-Witten superpotential}
\label{GVWW}

In this subsection, we make an important consistency check of our
gauge/gravity set up  by matching the GVW superpotential \cite{Gukov:1999ya} 
to the gauge theory effective superpotential considered in the
previous subsection.

It is well known that Calabi-Yau compactification of type IIB in the
presence of fluxes helps to restrict the allowed values of  the complex
structure moduli. The dynamics of these moduli can be encoded in an effective
superpotential $W_{GVW}$ for the resulting four  dimensional
supergravity. In the gauge/gravity correspondence
setup, $W_{GVW}$ can also be computed, provided  we fix some boundary
conditions at infinity on the non-compact CY we are using. It can be
written as 
\be
\label{GVWsuperppp} W_{GVW} = \frac{i}{2\pi g_s
\alpha'^4}\int_{\calM_6} G_3 \wedge \Omega ~, 
\ee 
where $\Omega$ is the holomorphic 3-form.  One can then compare this $W_{GVW}$
superpotential to the dual gauge theory superpotential, since they are
expected to agree on-shell.\footnote{Remark  that $W_{GVW}$ is a
supergravity superpotential, in particular $dW_{GVW}=0=W_{GVW}$ on
supersymmetric compactifications. To decouple gravity we must consider non-compact manifolds
and accordingly on the dual gauge theory side we only have
$dW=0$.}

In the absence of brane sources, the $G_3$ flux is closed and depends only on
the cohomology class of $G_3$. Adding some D5-brane sources for
$G_3$, however, one must keep track of the position  of these
branes \cite{Witten:1997ep,Aganagic:2000gs,Aganagic:2007py}. 
Separating $G_3$ into a bulk contribution (i.e. closed part)
$G_3^\text{b}$ and a contribution from the sources  $G_3^{s}$, and
using Riemann relations for the closed part, one has 
\be
\label{GVW explicitely} W_{GVW} = \frac{i}{2\pi g_s\alpha'^4} \sum_j
\Big( \int_{A_j} G_3^\text{b} \int_{B_j} \Omega - \int_{B_j}
(G_3^\text{b}+G_3^\text{s}) \int_{A_j} \Omega \Big) - \frac{2\pi
i}{\alpha'^3}\sum_\text{\Nugual{2} branes} \int_{\Xi_3} \Omega ~, 
\ee
where $\Xi_3$ is a 3-chain that extends from the 2-cycle wrapped by
the D5-brane to some reference 2-cycle near infinity.%
\footnote{For an intuitive feel for the meaning of that formula, one
can think of a one-dimensional analogy:
$dF^{(s)}=\delta_{\mathrm{source}}$ means that $F^{(s)}$ is a step
function that begins at the location of the source.  It is easy to
generalise the argument to 6 dimensions, at least formally by
integration by part.}

Let us now compute $W_{GVW}$ in our orbifolded conifold geometry. We
consider a generic smooth deformation, with the two complex  structure
parameters $\epsilon_1$, $\epsilon_3$ arbitrary, see \eqref{def orb conifold equation},
and we take the limit where the wrapped
D5-branes are far from the deformation near the tip.  With an obvious
linear change of coordinates, the geometry is defined by
\be
\label{Pdefoco} 
xy - (u^2 - v^2 + \epsilon_1)(u^2 - v^2 +
\epsilon_3) = 0~ 
\ee 
in $\bbC^4 \cong \{ x,y,u,v \}$.  The holomorphic
3-form $\Omega$ is given by 
\be 
\Omega = \frac{1}{2\pi^2} \,
\frac{du\wedge dv \wedge dx}{x} ~.  
\ee 
We obtain the usual results
for the periods of $\Omega$  on the $A$ and $B$ cycles (see appendices
\ref{sec: 3-cycles} and \ref{compute Omega periods} for more details)
\be 
\int_{A_j} \Omega = \epsilon_j, \qquad \text{and} \qquad
\int_{B_j} \Omega  =\  \frac{\epsilon_j}{2\pi i}  \log \left(
\frac{\epsilon_i}{4ev_0^2}\right) + \text{regular}~, 
\ee 
where $v = v_0$ is a cut-off for the non-compact $B$-cycles.  The contribution to
\eqref{GVW explicitely} coming from  D5-branes wrapped on $\calC_2$ is
computed in appendix \ref{compute Omega periods} : for a D5-brane
located at $v=\xi$, in the limit $|v_0|^2 \,,\, |\xi|^2 \gg
|\epsilon_k|$, we have the simple result 
\be
\label{simplePerXi3}
\int_{\Xi_3} \Omega = -\frac{1}{2\pi i} (\epsilon_1 - \epsilon_3) \log
\frac{\xi}{v_0} +\orders{\frac{\epsilon^2}{\xi^4}}~.  
\ee 
Let us now consider the following $F_3$ fluxes 
\be 
-\frac{1}{4\pi^2g_s
\alpha'}\int_{A_1} G_3^{b} = M_1, \qquad -\frac{1}{4\pi^2
g_s\alpha'}\int_{A_3} G_3^{b} = M_3-M_2~.  
\ee 
This means we assume that $M_1$ and $M_3-M_2$ D5-branes that were wrapped on the 2-cycles
$\calC_1$ and $\calC_3$, see eq.~\eqref{c1c3}, have undergone geometric
transition independently.%
\footnote{There is thus an arbitrariness in
choosing these fluxes, and we actually wrote the flux assignment that
makes the following arguments the simplest. The identifications (\ref{identi Btotau 1}) 
and (\ref{identi Btotau 2}) below consistently reflect this choice. }

We also have $M_2$ D5-branes wrapped on $\calC_2$, at positions
$|\xi_i|^2 \gg |\epsilon_{1,2}|$.  Let us finally denote the
$B$-periods of $G_3$ by the complex numbers 
\be
\label{BperiodsG3}
\calB_k \equiv - \frac{1}{4\pi^2 g_s\alpha'}\int_{B_k} G_3, \qquad  k=1,
\ 3~.  
\ee
Plugging all this into (\ref{GVW explicitely}) and denoting the product of the positions $\xi_i$ by $\xi^{M_2}$, 
we get
\be 
\alpha'^3 W_{GVW} =  -\epsilon_1
\ln{\left(\frac{\epsilon_1^{M_1}}{e^{M_1} (2v_0)^{2M_1-M_2}\
(2\xi)^{M_2}}e^{-2\pi i \calB_1} \right)} - \epsilon_3
\ln{\left(\frac{\epsilon_3^{M_3-M_2} \
(2\xi)^{M_2}}{e^{M_3-M_2}(2v_0)^{2M_3-M_2}}e^{-2\pi i \calB_3}
\right)}~.  
\ee
This flux plus branes configuration should correspond to the mesonic
branch of the gauge theory $(M_1,M_2,M_3,0)$ in the regime of section
\ref{regimeL2llL13}. In order to compare this superpotential to the
gauge theory result, we need to find the correct gauge/gravity
dictionary. Let us identify as usual the cutoff of the B-cycle
with the UV cutoff in the field theory, so that we  have
\be
\frac{1}{\alpha'^3} \,  (2v_0)^2 =   \mu_0^3 \qquad  \qquad \frac{1}{\alpha'^3} \
\epsilon_{1,3} = S_{1,3}~,
\ee
Naturally, $\mu_0$ is the UV scale at which we define the gauge
theory, while $S_1$ and $S_3$ are the gaugino condensates of  the first
and third node of the quiver.  We also know from the gauge theory
analysis that the eigenvalues $n_i$ of the meson matrix $\calN$ are to
be identified with the coordinates $z_2^{(i)}$ on the $p$-line of
singularities. More precisely
\be
n_i \ \propto \ z_2^{(i)} =
\xi_i+\sqrt{\xi_i^2+\epsilon} \ \approx \ 2\xi_i, \qquad \mathrm{for}
\quad |\xi_i|^2 \gg |\epsilon_{1,3}|~,
\ee
taking the root close to $\xi_i$. Equating the dimensionless
ratios $\xi/v_0 = n_i/\mu_0^2$ on both sides of
the correspondence, we find the relation 
\be 
\frac{1}{\alpha'^{3/2}}\
2\xi_i  \quad =\quad    \frac{n_i}{\mu_0^{1/2}}~.
\ee 
We still have to relate the B-periods of $G_3$ (\ref{BperiodsG3}) to
gauge theory quantities. This is the most subtle part, since these
periods are not topological, but instead depend crucially on the
boundary conditions at infinity (and hence on the bare Lagrangian of
the field theory).  By the non-renormalisation theorem, we know that
$W$ should not depend on the cut-off.  Imposing $\mu_0 \frac{\partial
W}{\partial \mu_0} =0$ gives us the following two conditions
\be
\label{constfrom RGinv} 
-2\pi i \frac{\partial \calB_1}{\partial \ln{
\mu_0}} = 3M_1 -2M_2 , \qquad \quad -2\pi i \frac{\partial
\calB_3}{\partial \ln{ \mu_0}} = 3M_3 -M_2 ~.  
\ee

In the particular case $M_1=0$, $S_1=0$, only the second condition has
to be imposed.  Then, it is easy to see that 
\be \label{identi Btotau 1}
\calB_3 \equiv \tau_0^{(3)} 
\ee 
should be identified with the UV value of the
holomorphic coupling of the third node, which provides the correct beta function. 
We have then reproduced the effective superpotential for the $(0,M_2,M_3,0)$ quiver, where
the second node is treated as a flavor group 
\be 
W =  - \ S_3\ln{\left(\frac{S_3^{M_3-M_2}\det{\calN}}{e^{M_3-M_2}
\mu_0^{3M_3-M_2} e^{2\pi i \tau_0^{(3)}}}  \right)} ~.  
\ee 
In the general case, $M_1\neq 0$, in order to satisfy the relations (\ref{constfrom RGinv}) 
we get for $\calB_1$ the identification 
\be \label{identi Btotau 2}
\calB_1 =
\tau_0^{(1)} +\frac{M_2}{2\pi i} \ln{(\mu_0 \lambda)}~, 
\ee 
where $1/\lambda$ is some scale, independent of $\mu_0$, that we will
identify with the inverse of the tree level quartic coupling in the gauge theory.

Defining the usual holomorphic SQCD scales 
\be 
\Lambda_1^{3M_1-M_2} =
\mu_0^{3M_1-M_2} e^{2\pi i \tau_0^{(1)}}, \qquad  \Lambda_3^{3M_3-M_2}
= \mu_0^{3M_3-M_2} e^{2\pi i \tau_0^{(3)}}~, 
\ee 
we then find the following superpotential
\be 
W =  M_1 S_1 \ - \ S_1
\ln{\left(\frac{S_1^{M_1-M_2}}{\Lambda_1^{3M_1-M_2}}\
\frac{S_1^{M_2}}{\lambda^{M_2}\det{\calN}} \right)} +
(M_3-M_2)S_3 \ - \
S_3\ln{\left(\frac{S_3^{M_3-M_2}\det{\calN}}{\Lambda_3^{3M_3-M_2}}
\right)}~.  
\ee 
This superpotential is precisely equal to the gauge theory result (\ref{superWMN13}), 
provided the first F-flatness condition of
(\ref{MeqNmin1}) is imposed.  This field theory constraint has a 
technical counterpart in our
analysis:  in supergravity we need to assume that a geometric
transition has taken place, so that we have a smooth geometry. Hence,
$\calC_2=\calC_4$ and the $p$- and $q$-lines meet smoothly, so there
is only one type of wrapped D5-brane to consider.  This is why we only
dealt with one single brane position $\xi$ while there are two
different mesons $\calM$ and $\calN$ in the gauge theory.

\subsection{IR regime and singularities resolution}

Let us now investigate how the backgrounds discussed in
section 3 must be modified at small radii in order to take into
account the non-trivial IR dynamics of the full physical quiver gauge
theory. As already mentioned previously, the dynamical scales at low
energies correspond to different dimensionful quantities in the
supergravity solution, depending on the qualitative dynamics of the
relevant node. For nodes 1 and 3, whose low-energy dynamics is
$\calN=1$, the scales $\Lambda_1, \Lambda_3$ are related to the
deformation parameters of the geometry $\epsilon_1, \epsilon_3$.  For node 2, which
leads essentially to $\calN =2$ dynamics, the scale  $\Lambda_2$ is
related to the enhan\c{c}on radius $\rho_c$ at which a probe $\calN=2$
fractional brane becomes tensionless; $\rho_c$ is related to the twisted 
flux terms in (\ref{flux N=2}).

Let us first briefly consider the regime where the dominant IR
dynamics is $\calN =2$, that is when $\Lambda_2 \gg \Lambda_1,
\Lambda_3$.  This translates in supergravity in a hierarchy where the
length scale defined by $\rho_c$ is much larger than the length scales
defined by $\epsilon_1=\epsilon_3$ (recall that the two deformation
parameters must be equal if there are BPS $\calN =2$ fractional branes
around). Since the enhan\c{c}on radius effectively cloaks
the singularity, length scales smaller than $\rho_c$ are not
accessible any more. Hence, the geometry which can be probed is always
at length scales for which the deformation is negligible. We thus
conclude that in this regime the UV solution of section 3
is a very good approximation even as far as the IR behavior is
concerned. Of course, the low-energy dynamics is $\calN=2$ in this
case and the gravity dual description of it has the usual drawback of
being essentially singular.

We now consider the richer case of the opposite regime, when
$\Lambda_1$, $\Lambda_3 \gg \Lambda_2$ and the dominant IR dynamics is
confining. Here we expect to be able to probe length scales where the
deformation drastically changes the underlying geometry.

There is actually a simple way to approach this problem. One can have 
BPS $\calN =2$ fractional branes in the
deformed geometry only when the two deformation parameters are
equal and the geometry is given by 
\be 
(z_1 z_2 - \epsilon)^2 = xy~.  
\ee 
As remarked in \cite{Argurio:2006ny}, this can
obviously be seen as the orbifold of the deformed conifold 
\be
z_1z_2-\epsilon =z_3z_4 
\ee 
under $\Theta : \,(z_1, z_2, z_3, z_4) \to (z_1, z_2,
-z_3,-z_4)$.  There is a single singularity line along $z_3=z_4=0\,,\, z_1z_2 =\epsilon$.

We can relate this complex form of the embedding to the real
coordinates on the deformed conifold as follows  
\bea
\label{coordinates def conif} 
z_1 &=  \sqrt{\epsilon}\,
e^{\frac{i}{2}(\phi_1 + \phi_2)} \Big\{ \sin\frac{\theta_1}{2}\,
\sin\frac{\theta_2}{2}\, e^{(\tau+i\psi)/2} + \cos\frac{\theta_1}{2}\,
\cos\frac{\theta_2}{2}\, e^{-(\tau+i\psi)/2} \Big\} \\ z_2 &=
\sqrt{\epsilon}\, e^{-\frac{i}{2}(\phi_1 + \phi_2)} \Big\{
\cos\frac{\theta_1}{2}\, \cos\frac{\theta_2}{2}\, e^{(\tau+i\psi)/2} +
\sin\frac{\theta_1}{2}\, \sin\frac{\theta_2}{2}\, e^{-(\tau+i\psi)/2}
\Big\} \\ z_3 &=  \sqrt{\epsilon}\, e^{-\frac{i}{2}(\phi_1 - \phi_2)}
\Big\{ \cos\frac{\theta_1}{2}\, \sin\frac{\theta_2}{2}\,
e^{(\tau+i\psi)/2} - \sin\frac{\theta_1}{2}\, \cos\frac{\theta_2}{2}\,
e^{-(\tau+i\psi)/2} \Big\} \\ z_4 &= \sqrt{\epsilon}\,
e^{\frac{i}{2}(\phi_1 - \phi_2)} \Big\{ \sin\frac{\theta_1}{2}\,
\cos\frac{\theta_2}{2}\, e^{(\tau+i\psi)/2} - \cos\frac{\theta_1}{2}\,
\sin\frac{\theta_2}{2}\, e^{-(\tau+i\psi)/2} \Big\} ~.
\eea 
Note that
$\tau$ is a dimensionless radial coordinate, and that for $\tau$
large $\epsilon \, e^{\tau} \rightarrow r^3$, we asymptote to the
singular conifold described in (\ref{coordC1}-\ref{coordC4}). We refer
to appendix \ref{sec: conifold} for the notation used hereafter.

The Calabi-Yau metric on the deformed conifold reads  
\ml{
ds_6^2 =
\frac{2^{\frac{2}{3}}}{3}  \epsilon^{2/3} K( \tau) \bigg[
\frac{1}{3K^3(\tau)} \Big( d\tau^2 + \zeta^2 \Big) +
\frac{1}{2}\sinh^2\frac{\tau}{2} \Big( (\sigma_1-\Sigma_1)^2 +
(\sigma_2-\Sigma_2)^2 \Big) \\
+ \frac{1}{2}\cosh^2\frac{\tau}{2} \Big(  (\sigma_1+\Sigma_1)^2 +
(\sigma_2+\Sigma_2)^2 \Big) \bigg] ~,
}
with
\be
K(\tau) = \frac{(\sinh \tau \cosh \tau- \tau)^{1/3}}{\sinh\tau}~.  
\ee
The orbifold action is  $(\phi_1,\phi_2) \to (\phi_1 - \pi, \phi_2 +
\pi)$, like in the singular case.  The fixed line at $z_3 = z_4 = 0$ is described 
by two halves:  $p = \{\theta_1 = \theta_2 = 0\}$ and $q = \{\theta_1 = \theta_2 = \pi\}$. 
We have 
\bea 
z_1\Big|_p &= \sqrt{\epsilon}\,
e^{-(\tau + i\, \psi')/2} \qquad\qquad & z_2\Big|_p &=
\sqrt{\epsilon}\, e^{(\tau + i\, \psi')/2} \\ z_1\Big|_q &=
\sqrt{\epsilon}\, e^{(\tau + i\, \psi'')/2} \qquad\qquad & z_2\Big|_q
&= \sqrt{\epsilon}\, e^{-(\tau + i\, \psi'')/2} 
\eea 
with $\psi' = \psi - \phi_1 - \phi_2$ and $\psi'' = \psi + \phi_1 + \phi_2$.
This line is completely smooth now: the $p$- and $q$-lines are glued together at $\tau=0$, with
the identification $\psi'=-\psi''$.  The full submanifold can
alternatively be described with a single patch, by extending the
domain of $\tau$  to $-\infty<\tau<+\infty$ and using, say, only
$\psi'$.  With this observation in mind, the metric on the
singularity line is  
\be
\label{cyltaupsi} ds^2 =
 \frac{2^{\frac{2}{3}}\epsilon^{2/3}}{9K^2(\tau)} \left(\rm{d}\tau^2
 +\rm{d}\psi'^2 \right)~.  
\ee 
It is a cylinder, on which we can introduce the complex coordinate $w= \tau+ i\psi'$.
We can construct the following 1-form on the line
\be 
\gamma=\frac{dz_2}{z_2}\Big|_p \equiv d\log z_2 =  \frac{1}{2} (d\tau + i\,
d\psi') = \frac{1}{2} dw~.  
\ee
Consider now a SUSY preserving ansatz similar to (\ref{ansatz}), but
with  a warped deformed conifold metric 
\be 
ds_{10}^2 = h^{-1/2}
dx_{3,1}^2 + h^{1/2} ds_6^2~.  
\ee 
The untwisted $G_3$ will be as in \cite{Klebanov:2000hb}, and the twisted 
part will get contribution by $\calN=2$ branes and by deformation branes, 
generically. It can be written as 
\ml{ 
G_3 = \frac{\alpha'}{2} g_s\, (-M_1+M_2-M_3) \Big[ \omega_3^{KS}
-\frac{i}{g_s} dB_2^{KS} \Big] \\ 
- 2\pi i\alpha' g_s
(M_1+M_2-M_3) \, d\log z_2 \wedge \omega_2 + 4\pi i\alpha' g_s
\sum_{j=0}^{M_2} d\log (z_2 - z_2^{(j)}) \wedge \omega_2
\label{def3form}
} 
where $\omega_3^{KS}$ and $dB_2^{KS}$ are the ones of
\cite{Klebanov:2000hb}.  In particular $d\omega_3^{KS} = 0$ and
$\int_{A_{CF}} \omega_3^{KS} = 8\pi^2$. Instead $\omega_2$ is the
anti-self-dual form at the orbifold point, normalised such that
$\int_{\calC_2} \omega_2 = 1$.  Moreover, $z_2^{(j)}$ are the
positions of the $M$ fractional branes on the $z_2$ plane. We get
\bea 
& -\frac{1}{4\pi^2\alpha'g_s} \int_{A_{CF}} F_3 = M_1-M_2+M_3 \\
& -\frac{1}{4\pi^2\alpha'g_s} \int_{A_2} F_3 = -M_1+M_2+M_3 \\ 
& -\frac{1}{4\pi^2\alpha'g_s} \int_{A_4} F_3 = M_1+M_2-M_3~, \\ 
\eea 
which exactly match those of the UV solution.
These integrals are easily performed by noticing that, in $A_2$, the
circle on the $p$ line at infinity is around $z_2 = \infty$, while in
$A_4\equiv - A_2$ the  circle on the $q$ line at infinity is around
$z_2= 0$.%
\footnote{Notice also that $A_2\cong \calC_2\times \psi'$  while $A_4
\cong \calC_4\times \psi''= -\calC_2\times \psi'$.}
The $M_2$ sources
provide for the difference between $\int_{A_2} F_3$ and $-\int_{A_4}
F_3$.

We can consider a simpler configuration, where the $\calN=2$
fractional branes are located at $\tau=\tau_0$ and are smeared on the
circle parametrized by $\psi'$.  We then consider 
\be 
\sum_{j=0}^{M_2} d\log (z_2 - z_2^{(j)}) \quad \rightarrow \quad  \frac{M_2}{2\pi i}
\oint \frac{dz_0}{z_0} d\log (z_2 - z_0)~, 
\ee  with $z_0=
\sqrt{\epsilon} e^{\frac{1}{2}(\tau_0 + \psi'_0)}$, and the integrand
is a differential in $z_2$.  The integral is thus performed at fixed
$\tau_0$.  It is easy to see that  
\be 
\frac{1}{2\pi i} \oint
\frac{dz_0}{z_0} d\log (z_2 - z_0) = \frac{dz_2}{z_2} \frac{1}{2\pi i}\oint
dz_0\big( \frac{1}{z_0} -\frac{1}{z_0-z_2} \big) ~.
\ee 
The integral is
vanishing if $|z_2|<|z_0|$ (that is $\tau<\tau_0$),  while it is unity
if $|z_2|>|z_0|$, which is $\tau>\tau_0$.  Hence, if we take the
branes to be smeared along the $\tau=0$ circle, the 3-form flux reads
\ml{
G_3 = \frac{\alpha'}{2} g_s (-M_1+M_2-M_3) \Big[ \omega_3^{KS}
-\frac{i}{g_s} dB_2^{KS} \Big] \\ - \pi i\alpha' g_s \Big[
(M_1+M_2-M_3)  - 2M_2\Theta (\tau) \Big] dw \wedge \omega_2 ~, 
}
where $\Theta $ is the Heaviside step function.  It is straightforward
to see that the twisted part of the 3-form
flux we get here is exactly equal to the one of the singular  conifold case
(\ref{flux N=2}).

The warp factor equation reads 
\be 
\Delta h = - \ast_6 (H_3\wedge F_3)
-\frac{M_2}{2} (4\pi^2\alpha')^2g_s \ \ast_6 \delta_6~.  
\ee 
We have included an explicit source term because in this case the source
branes are located at an otherwise smooth point of the geometry.  As
in the singular case, the twisted and untwisted 3-form terms do not
mix, and we can write the above equation in a way much similar to the
one appearing in (\ref{eq warp}). There will be a first, completely
smooth term on the r.h.s. coming from $\ast_6 (H_3^{KS}\wedge
F_3^{KS})$. The terms coming from the twisted flux will be similar to
the ones in (\ref{eq warp}), with a $\tau$-dependent prefactor.
Eventually, the term coming from the explicit source term will contain
a $\delta(\tau)$.  Of course, the warp factor will be a sum of the
particular inhomogeneous solutions of the Laplace equation with the
various source terms.  For instance, there will be a first piece which
will be given by  $h^{KS}(\tau)$. The other pieces will necessarily
involve a dependence on the other coordinates. Because of the
smearing, we can consider an ansatz for $h$ which does not depend on
$\phi_i$. However as we will see instantly, we will have to keep
explicit $\psi$ dependence in $h$.%
 \footnote{This is because  $\partial_{\psi}$ does not generate an
 isometry  of the deformed conifold. Hence smearing the sources
along $\psi$ does not help.}

The Laplacian on the deformed conifold  for
$h(\tau,\psi,\theta_1,\theta_2)$ reads (see also  the appendix of
\cite{Krishnan:2008gx}) 
\bea
\frac{2^{\frac{2}{3}}\epsilon^{\frac{2}{3}}}{3}\Delta h  = &
\frac{3}{\sinh^2\tau}\partial_\tau \big(K^2 \sinh^2\tau \partial_\tau
h) + 6 K^2 \partial_\psi^2 h\\ &  + \frac{2\cosh\tau}{K \sinh^2 \tau}
\Big( \partial_1^2 h +\cot \theta_1 \partial_1 h + \cot^2 \theta_1
\partial_\psi^2 h  + \partial_2^2 h +\cot \theta_2 \partial_2 h +
\cot^2 \theta_2 \partial_\psi^2 h \Big) \\ & + \frac{4}{K \sinh^2
\tau}\Big[ \cos\psi\big(\cot \theta_1\cot \theta_2 \partial_\psi^2 h -
\partial_1 \partial_2 h \big) +\sin\psi \big(\cot \theta_1
\partial_1\partial_\psi h +\cot \theta_2  \partial_\psi\partial_2
h\big) \Big]~. 
\label{deflapl} 
\eea
We see that the angular operator on
the third line has explicit dependence on $\psi$. A solution of the
Laplace equation independent on $\psi$ must then be also independent
of $\theta_1$ and $\theta_2$, which is not consistent with the
functional dependence of the source terms.  Hence we are forced to
consider a $\psi$ dependent warp function.

We can now view the Laplace operator on $h$ as a sum (weighted by
functions of $\tau$) of angular operators, which can be thought of as
acting on the variables defining the 5-dimensional space $T^{1,1}$.
The angular operators appearing in the first two lines are actually
the three angular operators which define the Laplacian on $T^{1,1}$,
$\partial_\psi^2$ and $ (\partial_i^2  +\cot \theta_i \partial_i  +
\cot^2 \theta_i \partial_\psi^2)$ for $i=1,2$, when they act on
functions which do not depend on the $\phi_i$ angles.  We can thus
find a complete basis of functions on $T^{1,1}$ which are
simultaneously eigenfunctions of these three operators.

In the deformed conifold however, we also have the additional angular
operator on the third line of eq.~(\ref{deflapl}). This operator will
inevitably mix eigenfunctions of the previous three operators, hence
making the problem of finding solutions to the Laplace equation a
problem of solving an (infinite) system of ordinary differential
equations.

Going over this analysis, even qualitatively or numerically, is
obviously beyond the scope of the present work. The main reason is
that locally,  the solution for the warp factor will again look like
the one for $\calN=2$ fractional branes at a $\bbC^2/\bbZ_2$
singularity, with its enhan\c{c}on-like singular behavior. Hence the
deep IR region has the difficulties common to the other $\calN=2$
gravity duals.  Nevertheless, it could be interesting to go further
along the analysis of the IR region of this configuration.

Let us now end this section with a very short remark on a particular
case, which is the one occuring when $M_2=0$. From the gauge theory
point of view, we expect a completely regular geometry 
\be
(z_1z_2-\epsilon_1)(z_1z_2-\epsilon_3)=xy ~.
\ee 
In particular, this
geometry no longer possesses lines of $A_1$-singularities.  However,
from the UV expression for the 3-form fluxes (\ref{flux N=2}) or
(\ref{def3form}), it seems that when $M_1 \neq M_3$ there is still a
twisted piece. This cannot be completely correct of course.
The $\epsilon_1 = \epsilon_3$ geometry is locally a $\bbC^2/\bbZ_2$ fibration
over the fixed line (topologically a cylinder).
When turning on different deformations
$\epsilon_1 \neq \epsilon_3$, the $\bbC^2/\bbZ_2$
singularity is blown-up fiberwise, with a base-dependent volume of the blown-up
2-cycle.
In particular its volume is a
$\tau$-dependent parameter
$a(\tau)$ such that $a \rightarrow 0$ when $\tau \rightarrow \pm
\infty$, while it reaches a maximum around $\tau=0$.
The 3-form can
be constructed from the ASD 2-form on the ALE space
which is the blow-up of $\bbC^2/\bbZ_2$, and is therefore
completely smooth in the bulk
of the geometry. However it asymptotes a $\delta$-function behaviour
for large radii, i.e. in the UV region. Hence, there is no
contradiction in the fact that the UV solution displays twisted flux also when there is no real orbifold fixed line.

\section{Discussion}
In this paper we have presented a supergravity solution which describes
fractional branes at the
orbifolded conifold.  The input is essentially given by the geometry
probed by the branes and its possible deformations, together with
the RR 3-form fluxes sourced by the fractional branes. The output can
be summarized in the NSNS 3-form flux and the warp factor, which 
should thus shed light on the characteristics of the dual gauge theory
which are not directly related to the holomorphic sector.

We have performed some non-trivial checks both on the UV
behavior of the NSNS flux, matching with a cascading interpretation of
the RG-flow of the gauge theory, and on the IR low-energy
theory by matching the effective superpotentials. The latter check  of
course only concerns the holomorphic sector, but clarifies the  IR
effects that the fractional branes have on the geometry.

The case where the supergravity solution is based on the deformed
geometry is the most interesting one. It corresponds to a hierarchy of
scales of the different nodes of the quiver that, in some specific
cases,  allows not only for supersymmetric vacua but also, possibly, for
metastable vacua \cite{Argurio:2006ny}.  The supersymmetric
supergravity solutions discussed here would correspond to the
supersymmetric vacua closest to the metastable vacua.  As was
suggested in \cite{Argurio:2006ny}, the gravity dual picture of the
metastable vacua, expected to be present for the gauge theory consisting
of 3-nodes with equal ranks $M$, is given in terms of $M$ anti-D3 branes in the
deformed geometry in the presence of 
$M$ fractional $\calN=2$ branes probing their moduli space.

As a first step towards the full supergravity description of the
metastable state, it would be interesting to consider the dynamics of an anti-D3 brane
probe in the geometry considered here (where possibly one would need
some more insight in the IR behavior of the warp factor).
In particular, it would be nice to see if the
supergravity solution indeed induces the expected attraction
towards the $\calN=2$ fractional D3-branes and
favors the anti-brane forming a bound state with them against the Myers
effect, 
which the anti-brane might undergo due to the presence of localized RR fluxes.

Notice however that in order to describe $M$ anti-D3 branes in a background where
the RR 3-form flux is also (exactly) given by $M$ units,  it is
necessary to consider their backreaction. Perturbatively, one can
perform a similar analysis as in \cite{DeWolfe:2008zy}. Incidentally,
it would be interesting to determine whether on the supergravity side the supersymmetry
breaking terms correspond in this case to F-terms on the field theory
side, as the analysis in \cite{Argurio:2006ny} suggests. The full
backreaction is clearly  a more ambitious goal. A possible avenue is
to consider what might be the endpoint of the interaction between the
anti-D3 branes and the wrapped D5-branes, namely a bound state where
the D3-charge of the wrapped branes has changed sign due to
supersymmetry breaking gauge flux on their world-volume. Possibly, the
latter picture is more amenable to a supergravity analysis along the
lines discussed in  this paper.


\vskip 10pt 
\centerline{\bf Acknowledgements}
\vskip 10pt 
\noindent
We would like to thank J.~Evslin, C.~Krishnan, S.~Kuperstein, M.~Mulligan, G.~Torroba 
and A.~Zaffaroni for helpful discussions.  This work is partially supported by the
European Commission FP6 Programme MRTN-CT-2004-005104, in which
R.A. and C.C. are associated to V.U. Brussel and F.B., M.B.  and
S.C. to Padua University.  R.A. is a Research Associate of the Fonds
de la Recherche Scientifique--F.N.R.S. (Belgium). C.C. is a Boursier
FRIA-FNRS.  The research of R.A. and C.C. is also supported by IISN -
Belgium (convention 4.4505.86) and by the ``Interuniversity Attraction
Poles Programme --Belgian Science Policy''.  S.C. is
grateful to the GGI Institute for Theoretical Physics in Florence for
ospitality  during the completion of this work.

\appendix

\section{Generalities on the conifold geometry}
\label{sec: conifold}

The singular conifold $C_0$ can be defined as an affine
variety in $\mathbb{C}^4 \cong \{z_1,z_2,z_3,z_4\}$,
\begin{equation}\label{con00}
z_1z_2 -z_3z_4=0~.
\end{equation}
By a linear change of coordinates, this can also be written as: $w_1^2 + w_2^2 + w_3^2 + w_4^2 = 0$.
The conifold is a CY cone, whose base is a Sasaki-Einstein
manifold called $T^{1,1}$ \cite{cdlo}. The latter is described
algebraically by the intersection of the cone with a unit
sphere in $\mathbb{C}^4$: $\sum_{i=1}^4 |w_i|^2 = 1$.
In terms of real coordinates, $w_i =
x_i + i y_i$, one gets $\vec{x} \cdot
\vec{x} = 1/2$, $\vec{y} \cdot \vec{y} = 1/2$, $\vec{x} \cdot
\vec{y} = 0$, which can be seen as an $S^2$ fibration over
$S^3$. However such a fibration is trivial%
\footnote{We can cover $S^3$
with two patches, intersecting at the equator. The bundle is
constructed by specifying a transition function on this equator
(itself an $S^2$), which is a map from $S^2$ to $SO(3)$, the structure
group of the fiber. Such maps are always trivial ($\pi_2(SO(3))=0$),
so the bundle is trivial.}%
, so that topologically $T^{1,1} \cong S^2\times S^3$.  The following coordinate system on the cone will be
useful%
\footnote{Remark that we differ from the conventions of
\cite{Klebanov:2000hb} by a flip in the orientation of the angles
$\phi_i$.}
\begin{align}
z_1 &= r^{3/2} \, e^{\frac{i}{2}(\psi+\phi_1+\phi_2)}
\sin\frac{\theta_1}{2} \sin\frac{\theta_2}{2}, \label{coordC1}\\ z_2
&= r^{3/2} \, e^{\frac{i}{2}(\psi-\phi_1-\phi_2)}
\cos\frac{\theta_1}{2} \cos\frac{\theta_2}{2}, \label{coordC2}\\ z_3
&= r^{3/2} \, e^{\frac{i}{2}(\psi-\phi_1+\phi_2)}
\cos\frac{\theta_1}{2} \sin\frac{\theta_2}{2}, \label{coordC3}\\ z_4
&= r^{3/2} \, e^{\frac{i}{2}(\psi+\phi_1-\phi_2)}
\sin\frac{\theta_1}{2} \cos\frac{\theta_2}{2} \label{coordC4}~.
\end{align}
Here,  $0 \leq \psi \leq 4 \pi~,~0 \leq \phi_i \leq 2 \pi~,~0 \leq
\theta_i \leq \pi$, and we have the following angular periodicities
\begin{equation}
\begin{pmatrix} \psi \\ \phi_1 \\ \phi_2 \end{pmatrix} \simeq \begin{pmatrix} \psi + 4\pi \\ \phi_1 \\ \phi_2 \end{pmatrix}
\simeq \begin{pmatrix} \psi + 2\pi \\ \phi_1 + 2\pi \\ \phi_2
\end{pmatrix} \simeq \begin{pmatrix} \psi + 2\pi \\ \phi_1 \\ \phi_2 +
2\pi \end{pmatrix}~.
\end{equation}

In these coordinates, the Calabi-Yau metric reads: $ds^2_{C_0} = \di r^2 +r^2\ ds^2_{T^{1,1}} $, with the Sasaki-Einstein metric of $T^{1,1}$
\begin{equation}\label{T11metr}
ds^2_{T^{1,1}} = \sum_{i=1,2} \frac{1}{6} \bigl( d\theta_i^2 +
\sin^2\theta_i \, d\phi_i^2 \bigr) + \frac{1}{9} \bigl( d\psi -
\sum_{i=1,1} \cos\theta_i \, d\phi_i \bigr)^2~.
\end{equation}
It describes a circle bundle, where the circle $\psi$ is fibered over
$S^2\times S^2$.
In terms of the natural vielbein for the two 2-spheres,
$u_i=d\theta_i$, $v_i=\sin\theta_i \,d\phi_i$ ($i=1,2$), it is useful
to define rotated vielbein for the 2-spheres \cite{Gwyn:2007qf}
\begin{equation}
\begin{pmatrix}\sigma_1\\ \sigma_2 \end{pmatrix}= \begin{pmatrix} \cos\frac{\psi}{2} 
& -\sin\frac{\psi}{2}\\ \sin\frac{\psi}{2} & \cos\frac{\psi}{2} \end{pmatrix}\begin{pmatrix} u_1\\v_1\end{pmatrix} \qquad\qquad
\begin{pmatrix}\Sigma_1\\ \Sigma_2 \end{pmatrix}= \begin{pmatrix} \cos\frac{\psi}{2} 
& -\sin\frac{\psi}{2}\\ \sin\frac{\psi}{2} & \cos\frac{\psi}{2} \end{pmatrix}\begin{pmatrix} u_2\\v_2\end{pmatrix} ~. 
\end{equation}
Let us also define $\zeta= d\psi - \sum_{i=1,2} \cos\theta_i \, d\phi_i$.  
For the singular conifold, we will use the following ordered vielbein
\begin{equation}\label{vielbein}
\left\{
e^r=dr\,,\;e^\psi=\frac{r}{3}\zeta\,,\;e^1=\frac{r}{\sqrt{6}}\sigma_1\,,\;
e^2=\frac{r}{\sqrt{6}}\sigma_2\,,\;
e^3=\frac{r}{\sqrt{6}}\Sigma_1\,,\;   e^4=\frac{r}{\sqrt{6}}\Sigma_2
\right\}~.
\end{equation}
The metric of the conifold then reads $ds^2_{C_0} = \sum_{n=1}^6 \, (e^n)^2$,
and the volume form is
\begin{equation}
\dvol_{C_0} = e^r\wedge e^\psi\wedge e^1\wedge e^2\wedge e^3\wedge e^4
= \frac{1}{108} \,r^5 \,dr \wedge d\psi\wedge  d\theta_1\wedge
\sin\theta_1\,d\phi_1 \wedge  d\theta_2\wedge \sin\theta_2\,d\phi_2~.
\end{equation}

A complex vielbein can be defined as
\begin{equation}
\left\{ E^1=e^1+ie^2\,,\;E^2=e^3+ie^4\,,\;E^3=e^r+ie^\psi\right\} ~.
\end{equation}
In terms of this complex structure, the K\"ahler form is
\begin{equation}\label{kahler}
J\equiv \frac{i}{2} \left( E^1\wedge \overline{E^1} +  E^2\wedge
\overline{E^2} +  E^3\wedge \overline{E^3}    \right) =
d\left(\frac{r^2}{6}\,\zeta\right)~,
\end{equation}
which is $(1,1)$, closed and satisfies $J\wedge J\wedge J= 6
\,\dvol_{C_0}$. It is exact, since we are at the zero resolution point
in K\"ahler moduli space where the cohomology class of $J$ is trivial.
The holomorphic top form is
\begin{equation}\label{Omega}
\Omega^{(3,0)}\equiv E^1\wedge E^2\wedge E^3 = -\frac{4}{9}
\, \frac{dz_1\wedge dz_2\wedge dz_3}{z_3}~.
\end{equation}

Let us now review 2- and 3-(co)cycles for the conifold.  We have the
closed (1,1)-form
\begin{equation}\label{omega2CF}
\begin{split}
\omega_2^{CF} & \equiv  \frac{3i}{2r^2}\left(E^1\wedge \overline{E^1}
- E^2\wedge \overline{E^2}  \right)=
\frac{1}{2}(\sigma_1\wedge\sigma_2 - \Sigma_1\wedge\Sigma_2 ) =\\ &=
\frac{1}{2}\left(\sin\theta_1 \,d\theta_1\wedge d\phi_1 - \sin\theta_2
\,d\theta_2\wedge d\phi_2 \right)~.
\end{split}
\end{equation}
The 2-cycle in $T^{1,1}$ is topologically a 2-sphere $\calC_{CF}$. It can be represented by
\begin{equation}
\calC_{CF}\,:\qquad \theta_1=\theta_2\equiv\theta\;,\quad
\phi_1=2\pi-\phi_2\equiv \phi\;,\quad \psi=0\;, \qquad
\phi\in[0,2\pi)\,,\;\theta\in(0,\pi)~.
\end{equation}
It turns out that $\int_{\calC_{CF}}\omega_2^{CF}=4\pi$.  In addition, one usually defines
the real closed 3-form
\begin{equation}\label{omega3CF}
\omega_3^{CF}\equiv \zeta\wedge \omega_2^{CF}~,
\end{equation}
which is the real part of the imaginary-self-dual (ISD) primitive
(2,1)-form
\begin{equation}\label{omega21}
\omega^{(2,1)}\equiv \frac{9}{2 r^3} E^3 \wedge \left(E^1\wedge
\overline{E^1} - E^2\wedge \overline{E^2}  \right) = \left( \zeta -3i
\frac{dr}{r}\right) \wedge \omega_2^{CF}~,
\end{equation}
defined on the whole conifold. Imaginary self-duality means that
$\ast_6 \,\omega^{(2,1)}= i\, \omega^{(2,1)}$.  The 3-cycle in
$T^{1,1}$ has the topology of a 3-sphere. We call it $A_{CF}$. It
can be represented by
\begin{equation}
A_{CF}\,:\qquad \theta_2=\phi_2=0~.
\end{equation}
Its orientation is such that
$\int_{A_{CF}}\omega_3^{CF}= 
8\pi^2$.

\section{The orbifolded conifold geometry}
\label{sec: orbconifold}
In this appendix, we derive the results presented in section
\ref{sec: gauge theory} concerning the relation between the ranks in
the quiver, the cycles wrapped by the different fractional branes, and
the fluxes present in the supergravity solution. In order to do this,
we need first to  discuss in detail the compact 2-cycles of the
geometry, on which the branes can wrap.  Then we discuss the compact
3-cycles of the geometry, which support the RR fluxes sourced by the
branes, and their intersections with the 2-cycles (in the base of the
singular cone). This will allow us to write the 3-form fluxes directly
in terms of the ranks of the gauge groups in the  quiver.

The CY singularity on which our gauge theory is engineered is a
non-chiral $\bbZ_2$ orbifold of  the conifold (\ref{con00}), obtained
considering the following action on the coordinates $z_i$ in $\bbC^4$
\be
\label{Z2 action on zapp}
\Theta\,: \quad (z_1, z_2, z_3, z_4) \;\to\; (z_1, z_2, -z_3, -z_4) ~.
\ee 
The orbifold geometry is still an algebraic variety. To describe
it one can introduce a complete set of invariants: $x \equiv z_3^2$,
$y \equiv z_4^2$ and $t \equiv z_3z_4$, which satisfy the constraint
$xy = t^2$. The conifold equation is rewritten as $t = z_1z_2$ so that
$t$ can be eliminated and we are left with 
\be
\label{orb conifold equation}
f = (z_1 z_2)^2 - x y = 0 ~.  
\ee 
The singular locus $f = df = 0$
consists of two complex lines that meet at the tip of the geometry
$\{z_1 = z_2 = x = y = 0 \}$, and corresponds to the fixed point locus
of the orbifold action $\Theta$.

One can use real coordinates as well, those already defined in
appendix \ref{sec: conifold}.  The orbifold action (\ref{Z2 action on
zapp}), which is an identification in the covering space, where we
will work, reads 
\be 
\Theta\,: \quad (\phi_1,\phi_2) \;\to\; (\phi_1 -
\pi, \phi_2 + \pi)~. 
\ee 
The two complex lines, that we call the $p$ and
$q$ line respectively, are defined, in complex and  real coordinates,
as 
\bea
\label{singularlines eq}
p &= \{ z_1 = x = y = 0,\, \forall z_2\} = \{ \theta_1 = \theta_2 =
0,\, \forall r, \psi' \} \\ q &= \{ z_2 = x = y = 0,\, \forall z_1 \}
= \{ \theta_1 = \theta_2 = \pi,\, \forall r, \psi'' \} ~, 
\eea where
$\psi'=\psi-\phi_1-\phi_2$ and $\psi''=\psi+\phi_1+\phi_2$ are (well
defined) angular coordinates along the singularity lines. In a
neighborhood of the singular lines (and outside the tip) the geometry
looks locally like the $A_1$-singularity $\bbC \times
\bbC^2/\bbZ_2$. The fixed point curve $p$ sits at the north poles of
both $S^2$'s while the curve $q$ sits at the south poles.

\subsection*{2-cycles and resolutions}
\label{sec: 2-cycles}

From the above analysis it follows that the singular geometry has
three vanishing 2-cycles.  Two of these three cycles arise due to the
orbifold action; such exceptional 2-cycles are located all along the
$\bbC^2/\bbZ_2$ singular lines $p$ and $q$ \eqref{singularlines eq},
and we call them $\calC_2$ and $\calC_4$, respectively.  Locally, one
could resolve the space into an
ALE space fibered over
$\bbC^*$. The third relevant 2-cycle descends from the 2-cycle of the
double covering conifold geometry,  whose base $T^{1,1}$ is topologically $S^2
\times S^3$.

Our goal in what follows is to pinpoint the precise map between
vanishing 2-cycles,  wrapped D5-branes,
3-form RR fluxes and quiver rank assignments. To this end, it  will
prove useful to take advantage of our CY cone being a toric
variety\footnote{A toric manifold is a  manifold of complex dimension
$r$ which admits an isometry group (at least as big as) $U(1)^r$. A
toric CY threefold is then  a CY threefold whose isometry group is at
least $U(1)^3$. For a recent introduction, see e.g. 
\cite{Denef:2008wq}.}, since in this case  one can use standard techniques
to understand the structure of 2-cycles and their intersections.  Let
us sketch how this comes about.

A toric variety can be described as the moduli space of an  associated
supersymmetric gauged linear $\sigma$-model (GLSM). Consider $n$
chiral superfields $t_i$, $i=1\ldots n$ charged under a product of
abelian gauge groups  $U(1)^s$, with charges $Q\du{a}{i}$, $a= 1
\ldots s$. In the absence of a superpotential,  the potential for the
scalar components is 
\be V(t_i) = \sum_{a=1}^s \Big( \sum_{i=1}^n
Q\du{a}{i} \, |t_i|^2 - \xi_a \Big)^2 ~.  
\ee 
where $\xi_a$ are
Fayet-Iliopoulos parameters (FI). The moduli space of vacua $\calM$ is
given by the D-flatness equations modulo $U(1)^s$ gauge
transformations 
\be \calM = \Big\{ t_i \in \bbC^n \Big| \sum_{i=1}^n
Q\du{a}{i} \, |t_i|^2 = \xi_a \quad \forall a=1,\dots,s \Big\} \Big/
U(1)^s ~, 
\ee 
where $U(1)^s$ acts as $t_i \to e^{i\, Q\du{a}{i}
\,\phi^a} t_i$. When the FI's are such that $\dim \calM = n-s$,
$\calM$ is the desired toric variety (and $n-s=r$ is just the number of
isometry abelian factors).  Putting the FI's to zero the variety, if
admissible, is scale  invariant: this corresponds to a cone. As the
FI's change, the K\"ahler moduli of $\calM$ also change and one gets
resolutions or blow-ups. Generically, different regions in the
parameter space of the FI parameters correspond to different
resolutions, delimited by flop transition curves.

In our case the GLSM has six fields $t_i$ whose charges $Q\du{a}{i}$
are reported in the table below 
\be
\label{table GLSM charges}
\begin{tabular}{cccccc|l}
$t_1$ & $t_2$ & $t_3$ & $t_4$ & $t_5$ & $t_6$ \\ \hline 0 & 0 & 1 &
$-2$ & 1 & 0 & $\xi_2$ \\ 1 & $-1$ & 0 & 1 & $-1$ & 0 & $\xi_\beta$ \\
$-2$ & 1 & 0 & 0 & 0 & 1 & $\xi_4$
\end{tabular}
\ee 
We can parameterize the toric variety with the gauge invariants
\be 
t_3t_4t_5 = z_1 \qquad t_1t_2t_6 = z_2 \qquad t_1t_2^2t_3^2t_4 = x
\qquad t_1t_4t_5^2t_6^2 = y 
\ee 
which, consistently, satisfy the
defining equation \eqref{orb conifold equation}. We can also give a
parametrization for the so-called toric divisors, which are the
four-dimensional hypersurfaces in the toric  CY defined by $D_i = \{
t_i = 0 \}$. We recognize $D_4 = \{z_1 = x = y = 0\}$ as the $p$ line
and $D_1 = \{ z_2 = x = y = 0 \}$ as the $q$-line.

\begin{figure}[tn]
\begin{center}
\includegraphics[height=3.5cm]{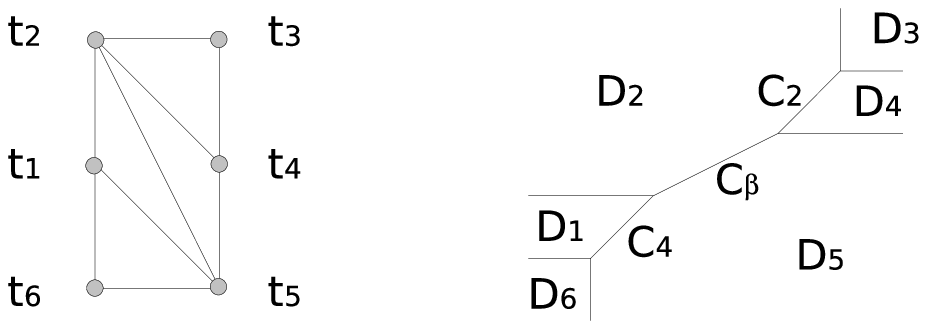}
\caption{\small The toric diagram and the dual $(p,q)$-web. The
specific toric diagram triangulation is the one related   to having
all $\xi_a>0$ in the associated GLSM.
\label{fig: toric orb conifold}}
\end{center}
\end{figure}

The toric diagram and the related $(p,q)$-web corresponding to
choosing all $\xi_a>0$ (which amounts to a given  triangulation of the
toric diagram) are depicted in Figure \ref{fig: toric orb conifold}.
For the particular resolution corresponding to $\xi_2, \xi_\beta,
\xi_4 >0$ the three holomorphic 2-cycles can be directly read from the
$(p,q)$-web. They can be explicitly constructed as intersections of
toric divisors 
\be 
\calC_2 = D_2 \cdot D_4 \qquad\qquad \calC_\beta =
D_2 \cdot D_5 \qquad\qquad \calC_4 = D_1 \cdot D_5 ~.  \ee This can be
explicitly checked using D-term equations, which for the intersections
of interest are 
\bea 
&D_2D_4 \,: \quad |t_3|^2 + |t_5|^2 = \xi_2
\qquad |t_6|^2 = 2|t_1|^2 + \xi_4 \qquad |t_1|^2 = |t_5|^2 + \xi_\beta
\\ &D_2D_5 \,: \quad |t_4|^2 + |t_1|^2 = \xi_\beta \qquad |t_3|^2 =
2|t_4|^2 + \xi_2 \qquad |t_6|^2 = 2|t_1|^2 + \xi_4 \\ &D_1D_5 \,:
\quad |t_2|^2 + |t_6|^2 = \xi_4 \qquad |t_3|^2 = 2|t_4|^2 + \xi_2
\qquad |t_4|^2 = |t_2|^2 + \xi_\beta ~.  
\eea 
As one can see, each
$\calC_i$ topologically is a $\bbP^1$ (parameterized by the first two
variables in each row) of volume $\xi_i$.

Let us consider also  another basis of 2-cycles, which arises in a
different resolution of the singular conical geometry (corresponding
to a different triangulation of the toric diagram).  Consider the
region in the space of FI parameters where $\xi_\beta <0$ with  $\xi_2
+ \xi_\beta>0$ and $\xi_4 + \xi_\beta >0$. We can introduce 
\be 
\xi_1
= \xi_4 + \xi_\beta >0 \qquad\qquad \xi_3 = \xi_2 + \xi_\beta >0
\qquad\qquad \xi_\alpha = - \xi_\beta >0~.  
\ee 
This new resolution
can be obtained from the one in Figure \ref{fig: toric orb conifold}
with a flop transition on $\calC_\beta \leftrightarrow
\calC_\alpha$. The toric diagram triangulation and the corresponding
dual $(p,q)$-web for the new geometry are sketched in Figure \ref{fig:
toric orb conifold flopped}. In order to have a nice presentation of
the GLSM charges in terms of the new positive FI's, we can linearly
re-shuffle Table \eqref{table GLSM charges} getting 
\be
\label{table
GLSM charges flopped}
\begin{tabular}{cccccc|l}
$t_1$ & $t_2$ & $t_3$ & $t_4$ & $t_5$ & $t_6$ \\ \hline $-1$ & 0 & 0 &
1 & $-1$ & 1 & $\xi_1$ \\ $-1$ & 1 & 0 & $-1$ & 1 & 0 & $\xi_\alpha$
\\ 1 & $-1$ & 1 & $-1$ & 0 & 0 & $\xi_3$
\end{tabular}
\ee
\begin{figure}[tn]
\begin{center}
\includegraphics[height=3.5cm]{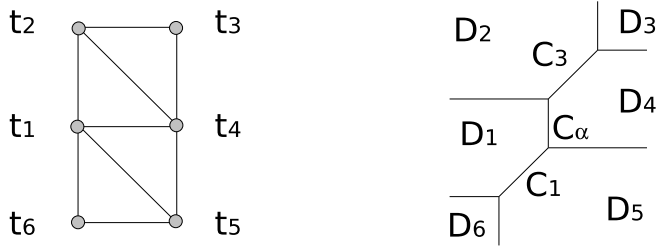}
\caption{\small The toric diagram and the dual $(p,q)$-web in the
region of the FI parameter space where $\xi_\beta <0$.
\label{fig: toric orb conifold flopped}}
\end{center}
\end{figure}
Repeating the same analysis as before one finds the holomorphic%
\footnote{Notice that generically if an homology class $\calC$ has a
holomorphic representative, $-\calC$ does not because the
representative becomes antiholomorphic and one should look for a
different one. In particular, in different resolutions the r\^ole of
homology classes with a holomorphic representative is exchanged.}
2-cycles in this new resolution in terms of toric divisors 
\be 
\calC_3
= D_2 \cdot D_4 \qquad\qquad \calC_\alpha = D_1 \cdot D_4 \qquad\qquad
\calC_1 = D_1 \cdot D_5 ~.  
\ee 
Again the FI parameters are the
positive volumes of the corresponding 2-cycles $\calC_i$. From the
relations among FI parameters we read the relations  
\be 
\label{c1c3}
\calC_1 = \calC_4 + \calC_\beta \qquad\qquad \calC_3 = \calC_2 + \calC_\beta ~,
\ee 
which can be thought of as relations in homology between vanishing
cycles.

A comment is in order at this point. In this non-chiral case, vanishing 2-cycles are in
one-to-one correspondence with possible fractional branes. All the
divisors  are non compact 4-cycles. This implies that all dual
2-cycles support non-anomalous fractional branes. This does not hold
in general, as only 2-cycles dual to non-compact 4-cycles give
anomaly-free fractional branes, their number being equal to the number
of 3-cycles in the real base of the CY cone (which in turn corresponds
to the number of baryonic charges). This is the geometric counterpart
of the dual gauge theory being non-chiral. Conversely, chiral theories
are related to CY cones where  there are compact 4-cycles around. The
latter put constraints on the allowed fractional D3-branes
configurations, because of the RR tadpole cancellation condition.

Once we wrap a D5-brane on a 2-cycle, it will thus source a 3-form RR
flux.  We turn to consider the compact 3-cycles of the
geometry which can  support this flux, and their dual non-compact
3-cycles.

\subsection*{3-cycles and deformations}
\label{sec: 3-cycles}

The study of compact and non-compact 3-cycles is best performed in a
regular geometry obtained by complex deformation of the singular
space, rather than by resolution (which is a K\"ahler deformation).

The algebraic variety \eqref{orb conifold equation} admits two
normalizable complex deformations parameterized by $\epsilon_1$ and
$\epsilon_3$ \cite{Argurio:2006ny} 
\be
\label{def orb conifold equation} f = (z_1z_2 -
\epsilon_1)(z_1 z_2 - \epsilon_3) - xy = 0 ~.  
\ee 
The deformed
geometry is regular for $\epsilon_1 \neq \epsilon_3$, provided
$\epsilon_1 \epsilon_3\neq 0$. For $\epsilon_1 = \epsilon_3\neq
0$ it still has a $\bbC^*$ line of $A_1$ singularities (locally $\bbC
\times \bbC^2/\bbZ_2$) and corresponds to a $\bbZ_2$ orbifold of the
deformed conifold. For $\epsilon_3=0$ it has a conifold singularity at
the tip.

\begin{figure}[tn]
\begin{center}
\includegraphics[height=6cm]{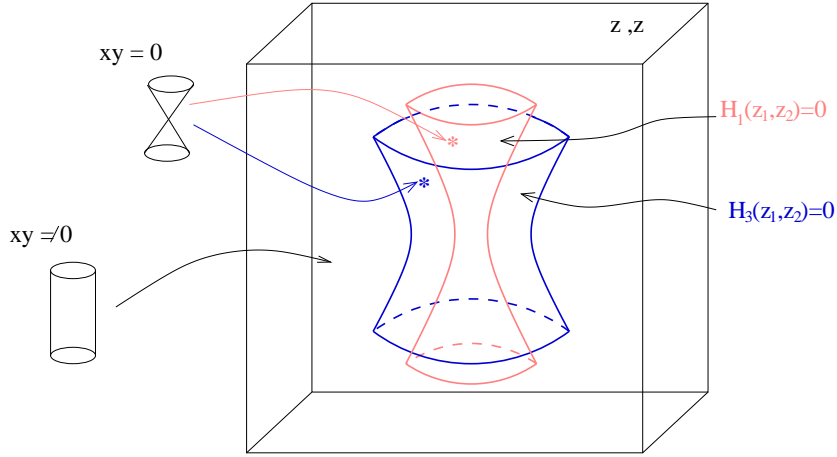}
\caption{\small The 6-dimensional manifold seen as a singular $\bbC^*$
fibration over the $(z_1,z_2)$ space. The surfaces $H_k(z_1,z_2)=z_1
z_2 - \epsilon_k=0$, $k=1,3$, are the loci where the $\bbC^*$ fiber
degenerates to a cone $xy=0$ and a non-trivial $S^1$
shrinks. \label{fig: deformed geometry}}
\end{center}
\end{figure}

A convenient way to visualize the geometry is to regard \eqref{def orb
conifold equation} as a singular $\bbC^*$ fibration over $\bbC^2
\simeq (z_1,z_2)$ 
\be xy = H_1(z_1,z_2) \, H_3(z_1,z_2) \qquad\qquad
\text{with}\qquad H_k (z_1,z_2) = z_1 z_2 - \epsilon_k ~.   
\ee 
At any point $(z_1,z_2)$ where $H_1(z_1,z_2) H_3(z_1,z_2) \neq 0$ the
fiber has equation $xy=c \neq 0$ and is a copy of $\bbC^*$. On each
surface $H_k(z_1,z_2)=0$ the fiber degenerates to a cone $xy=0$ and an
$S^1$ shrinks. On the other hand, each surface $H_k(z_1,z_2)=0$ is an
hyperboloid in $\bbC^2$ and has the topology of $\bbC^*$. For a
general deformation, $\epsilon_1 \neq \epsilon_3$, they are disjoint
and never touch. When $\epsilon_1 = \epsilon_3$ they degenerate one on
top of the other, while when one deformation parameter vanishes the
corresponding hyperboloid degenerates into a cone. See Figure
\ref{fig: deformed geometry} for a picture of the geometry.

Figure \ref{fig: deformed geometry} is very useful to visualize
compact and non-compact 3-cycles as well as 2-cycles in the deformed
geometry. Any line segment of real dimension one in the $\bbC^2$ space
$(z_1,z_2)$ which begins and ends on the locus $xy=0$ represents a
closed submanifold of real dimension two, obtained by fibering on that
segment an $S^1$ which lives in the $\bbC^*_{x,y}$ cylinder and
shrinks to zero at the endpoints. When the line segment is
non-contractible (keeping the endpoints on the $xy=0$ locus), it
represents a non-trivial element in the homology group
$H_2(\calM,\bbZ)$. In the same way, a real dimension two surface with
boundary on the $xy=0$ locus gives rise to a closed dimension three
submanifold after the $S^1$ has been fibered on it. When the surface
is non-contractible (keeping the boundary on the $xy=0$ locus),  it
gives rise to a non-trivial 3-cycle. Compact 3-cycles $A_i$ arise from
compact surfaces while non-compact 3-cycles $B_i$ arise from
non-compact surfaces.

\begin{figure}[tn]
\begin{center}
\includegraphics[height=6cm]{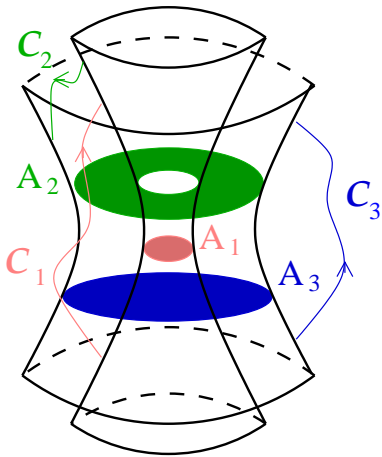}
\caption{\small The projection of the $A$ and ${\cal C}$ cycles in the
$(x,y)$ space. The non-compact B-cycles are obtained as ${\cal C}$-cycles
fibers over $r$. \label{fig: def geometry cycles}}
\end{center}
\end{figure}
In Figure \ref{fig: def geometry cycles} we depicted the various
2-cycles $\calC_i$ and compact 3-cycles $A_i$ for the deformed
orbifolded conifold. We have used the basis which is most natural when
complex deformations are concerned. Non-compact 3-cycles $B_i$ are
easily obtained as well: the real dimension two base surfaces are
non-compact ``vertical'' foils with one or two boundaries on the
degeneration loci, and are related to the line segment supporting the
2-cycles $\calC_i$.

In the regular deformed geometry, a canonical symplectic basis for the
third homology group $H_3(\calM, \bbZ)$ is given by $\{A_1, A_3, B_1,
B_3\}$ with intersection numbers $A_i \cdot B_j = \delta_{ij}$.
 $A_1$ and $A_3$ have topology $S^3$ while
$B_1$ and $B_3$ have topology $\bbR^3$. One can also consider a linear
combination of them, $A_2 = A_1 - A_3$ (see Figure \ref{fig: def
geometry cycles})  and its dual $B_2 = -B_1 + B_3$ :  they have
intersection number $A_2 \cdot B_2 = -2$.

The asymptotic behavior of supergravity solutions based on these
spaces is fixed, among other parameters, by the D5-charges at
infinity. These are constructed by integrating suitable currents on
the 3-cycles in radial sections of the asymptotically conical
geometry. This is equivalent to considering any radial section in the
singular conical geometry ($\epsilon_1 = \epsilon_3 = 0$). The latter
perspective is useful because from any 3-cycle in a radial section we
can construct a non-compact conical 4-cycle having the 3-cycle as its
radial section: this allows us to introduce a concept of holomorphy
and to use toric divisors instead of 3-cycles in radial sections.

From the GLSM description we know that the number of compact 3-cycles
in radial sections (which equals the number of baryonic charges and
the number of non-anomalous fractional branes) is three. For
concreteness we choose the following basis: $A_2$, $A_4$ and
$A_{CF}$. $A_2$ is the radial section of the toric divisor $D_4$, and
corresponds to the product of the exceptional 2-cycle $\calC_2$ along
the $p$-line (which is $\cong \bbC^*$) with  $S^1$ in the latter; in
the same way, $A_4$ is the radial section of the toric divisor $D_1$,
and is the product of the exceptional $\calC_4$ along the $q$-line
times $S^1$. $A_{CF}$ is the compact 3-cycle of the covering space
conifold\footnote{Actually $A_{CF}=A_1+A_3$.}:  under the orbifold
action it has an image, and no fixed points.  In particular, the
representative 3-cycle at $\theta_2 = \pi/2$ and $\phi_2 = 0$ is
mapped to the divisor $\{ x=z_1^2, y=z_2^2 \}$ which has the GLSM
description $t_1t_2^2 = t_4 t_5^2$. Comparing the charges we find that
$A_{CF}$ corresponds to the toric divisor $D_1 + 2D_2 = D_4 +
2D_5$. Summarizing, our basis of 3-cycles and the corresponding toric
divisors are 
\be 
A_2 \simeq D_4 \qquad\qquad A_4 \simeq D_1
\qquad\qquad A_{CF} \simeq D_1 + 2D_2 = D_4 + 2D_5 ~.   
\ee  
Notice
that in the deformed geometry $A_2 = -A_4$ in homology. Nevertheless
they can give rise to different charges when explicit sources are
present in the geometry and this is in fact the case of \Nugual{2}
branes which do not undergo complete geometric transition.

In order to compute the 3-form fluxes  generated by D5-branes wrapped
on 2-cycles, we will need the intersection matrix between divisors and
2-cycles. In our basis we find  
\be
\label{table intersection numbers}
\begin{tabular}{l|ccc}
 & $A_2 \simeq D_4$ & $A_4 \simeq D_1$ & $A_{CF}$ \\ \hline $\calC_2$
& $-2$ & 0 & 0 \\ $\calC_4$ & 0 & $-2$ & 0 \\ $\calC_\beta$ & 1 & 1 & $-1$
\end{tabular}
\ee  
This table is computed from the charges in Table \eqref{table
GLSM charges}: in the GLSM construction each gauge field gives rise to
an element $\calC_a$ of the homology group $H_2(\calM,\bbZ)$, and the
intersection between it and a toric divisor $D_i$ is the charge
$Q\du{a}{i}$.%

\subsection*{The fractional branes/ranks correspondence}

We have now all the ingredients to finally figure out the precise
correspondence  between fractional branes (that is wrapped D5-branes)
and quiver rank assignments.

Consider a D5-brane wrapped on a 2-cycle $\calC_i$ of our CY$_3$. The
Bianchi identity for $F_3$ is violated by the source 
\be 
dF_3 = -2\kappa^2 \tau_5 \, \Omega_4 ~, 
\ee 
where $\Omega_4$ is a 4-form with
$\delta$-function support on the D5 world-volume.  We are interested
in the flux generated on a 3-cycle $A_j$ in the radial section. First
we have to resolve the geometry, switching on the FI parameters of the
associated GLSM. This does not change the holomorphic data nor the
quantized charges. Then we identify a non-compact divisor $D_j$ which
has $A_j$ as radial section. Being the geometry smooth, $A_j$ turns
out to be the boundary of $D_j$  
\be
\label{def orient 3cycles}
\int_{A_j} F_3 = -\int_{D_j} dF_3 = 2\kappa^2 \tau_5 \int_{D_j}
\Omega_4 = 2\kappa^2 \tau_5 \: (D_j \,,\, \calC_i) ~, 
\ee 
where $(D_j\,,\,\calC_i)$ is the intersection number as in Table
\eqref{table intersection numbers}, and we fixed the orientation ambiguity requiring consistency with known cases, such as the conifold and the $\bbZ_2$ orbifold of $\bbR^6$.  If there is a holomorphic
representative for $\calC_i$, we can then directly compute the
intersection from the GLSM data.

The last thing to determine are the quiver rank assignments
corresponding to each fractional brane. A D5-brane wrapped on the
exceptional 2-cycles $\calC_2$ and $\calC_4$ along the $\bbC^2/\bbZ_2$
lines $p$ and $q$ gives rise to an \Nugual{2} fractional brane, and we
conventionally choose the rank assignments to be, respectively,
$(0,1,1,0)$ and $(1,1,0,0)$. The rank assignment for a D5-brane
wrapped on $\calC_\beta$ can be defined by observing that the
combination $\calC_{CF} = 2\calC_\beta + \calC_2 + \calC_4$ does not couple to
twisted fields and gives rise to the orbifold of the Klebanov-Tseytlin
theory \cite{Klebanov:2000nc}, see Table (\ref{table intersection
numbers}).  This implies that the corresponding gauge theory is the
orbifold of the KT theory. We can say that the ranks for one D5 on
$C_\beta$ are $(a,b,c,d)$. Requiring that $2C_\beta + C_2 + C_4$ is in
the class $(N+1, N, N+1,N)$ or $(N, N+1, N,N+1)$, which do correspond
to the orbifold of the KT theory,  singles out two possibilities for
$C_\beta$: either $(1,0,1,1)$ or $(0,0,0,1)$. To select the correct option we should consider  the
induced D3-charge on the fractional D3 probe.

The induced D3-charge is
proportional to the integral of $B_2$ (or more generally of $\calF = B_2 + 2\pi\alpha' F_2$) on the
corresponding 2-cycle $\calC$:
\be 
Q_3 \ =\  \tau_5 ~ \int_{\calC} \calF \ =\
\tau_3~\frac{1}{4\pi^2\alpha'} \int_{\calC} \left(B_2 + 2\pi\alpha'
F\right) ~.
\ee 
The actual value depends on the background value of $B_2$. This is arbitrary at this level (and it is
related to the UV cut-off values of the gauge couplings in the dual
gauge theory). We  only require these  background values to be
positive (so as to describe mutually BPS objects) and less than one
(in order to describe non-composite, that is elementary, objects).
Along the $p$ and $q$ lines the physics is locally $\bbC^2/\bbZ_2$,
thus we can naturally set \cite{Douglas:1996sw}: $\int_{C_2} B_2 = \int_{C_4} B_2 = (4\pi^2\alpha')/2$. If we consider the KT theory and set also \cite{Klebanov:2000nc} 
$\int_{\calC_{CF}} B_2 = (4\pi^2\alpha')/2$, then using the previous relation $\calC_{CF} = 2C_\beta + C_2 + C_4$, we get $\int_{C_\beta} B_2 = -(4\pi^2\alpha')/4$.

This implies that while the \Nugual{2} branes have positive
D3-charge, a D5-brane wrapped on $C_\beta$ has  negative D3-brane
charge and it is not mutually BPS. Putting one unit of
worldvolume flux on the wrapped D5 we get positive D3-charge: $3/4$. The total D3-charge for $\calC_{CF} = 2C_\beta + C_2 + C_4$ (with two units
of flux on $C_\beta$) is $5/2$. This is exactly the D3-charge of the
configuration $(3,2,3,2)$, which implies that one D5-brane wrapped on
$C_\beta$ with one unit of worldvolume flux gives rise to the theory
$(1,0,1,1)$.  A similar analysis shows that a D5-brane wrapped on
$\calC_\alpha=-\calC_\beta$ (with no background world-volume flux)
corresponds to a rank assignement $(0,1,0,0)$.  Finally, direct
application of Table (\ref{table intersection numbers}) tells us  what
the fluxes sourced by D5-branes wrapped on any 2-cycles are.

Our findings are summarized in the Table below 
\be
\label{table summary frac branes bis}
\begin{tabular}{l|ccccc}
 & $-\int_{A_2} F_3$ & $-\int_{A_4} F_3$ & $-\int_{A_{CF}} F_3$ &
D3-charge & gauge theory \\ \hline D5 on $\calC_2$ & 2 & 0 & 0 &
1/2 &(0,1,1,0) \\ D5 on $\calC_4$ & 0 & 2 & 0 & 1/2 & (1,1,0,0) \\
D5 on $\calC_\beta$ & $-1$ & $-1$ & $1$ & 3/4 & (1,0,1,1)\\ D5 on
$\calC_\alpha$ & $1$ & $1$ & $-1$ & 1/4 & (0,1,0,0)
\end{tabular}
\ee  
where fluxes are in units of $4\pi^2\alpha'g_s$.

As anticipated, we will use D5 branes wrapped on $\calC_2$, $\calC_4$ and
$\calC_\alpha=-\calC_\beta$ without worldvolume flux as a basis
for fractional branes to discuss our gauge/gravity duality. This is
the most natural basis for discussing rank  assignments parametrized
as in Figure \ref{cz2}, where fractional branes modify the ranks of
the first three quiver nodes only.

\section{Conventions: action, charges and EoM}
\label{sec: action}

We follow conventions according to which the action of Type IIB
supergravity reads, in Einstein frame
\ml{ \label{sugraIIB}
S_{IIB} = \frac{1}{2\kappa^2} \biggl\{ \int d^{10}x \, \sqrt{-g} \, R
\; -\frac{1}{2} \int \Bigl[ d\Phi\wedge\ast d\Phi + e^{2\Phi}
F_1\wedge\ast F_1 + \frac{1}{2} F_5\wedge\ast F_5 \\
+ e^{-\Phi} H_3\wedge\ast H_3 +
e^\Phi F_3\wedge\ast F_3 - C_4\wedge H_3 \wedge F_3 \Bigr] \biggr\} ~,
}
where $\kappa^2= \pi (2\pi)^6 \alpha'^4 g_s^2$ is the Newton coupling
constant and the gauge invariant field strengths are defined as
\begin{equation}
F_1=dC_0\;,\qquad F_3=dC_2+C_0 H_3\;,\qquad F_5=dC_4+C_2\wedge
H_3\;,\qquad H_3=dB_2 ~.
\end{equation}
In our conventions the Einstein frame is defined from the string frame
by rescaling the metric by the {\it fluctuating} part of the dilaton
field. Moreover, our RR fields are normalized so as to 
appear in the action in a democratic way with respect to the NSNS fields,
that is the Newton coupling constant $\kappa$ enters as an overall
factor in front of the Einstein frame supergravity action. As a
consequence, the dilaton field $\Phi$ appearing in the action
(\ref{sugraIIB}) is its fluctuating part, only, as its VEV has been
absorbed into $\kappa$. With these conventions, the world-volume
action for a Dp-brane is
\begin{equation}
S^{Dp}_{loc} = - \tau_p \int_{Dp} d^{p+1}\xi \, e^{\frac{p-3}{4}\Phi} \,
\sqrt{-\det(\hat g+ e^{-\Phi/2}\, \calF)} + \tau_p \int C \wedge
e^\calF \wedge \Omega_{9-p} ~,
\label{wvdp}
\end{equation}
where $\calF = \hat B_2 + 2\pi\alpha'\, F_2$ (the hat on the NSNS
2-form means that the form is pulled-back on the D-brane world-volume)
and $\tau_p = 1/[(2\pi)^{p} \alpha'^{\frac{p+1}{2}}g_s]$. Finally, $C$
is a polyform $C = \sum C_p$, with $C_p$ being all possible RR
potentials, and $\Omega_{9-p}$ is a form localized on the Dp-brane
worldvolume (the Poincar\'e dual to the cycle) and closed.

With these conventions, the D3-brane and D5-brane (Maxwell) charges
are, respectively
\begin{equation}
\label{maxcharge}
Q_{D3}= -\frac{1}{(4\pi^2\alpha')^2\,g_s} \int F_5\qquad,\qquad
Q_{D5}= -\frac{1}{4\pi^2\alpha'\,g_s} \int F_3  ~.
\end{equation}
The equations of motion for the fields relevant to our solution are
\begin{equation} 
\begin{aligned}
d ~e^\phi \ast F_3 &= H_3 \wedge F_5 - 2 \kappa^2 \frac{\delta
S_{loc}}{\delta C_2}  \\ d F_5 &= - H_3 \wedge F_3 - 2 \kappa^2
\frac{\delta S_{loc}}{\delta C_4} \\ d ~e^{-\phi} \ast H_3 &= - F_3
\wedge F_5 - 2\kappa^2 \frac{\delta S_{loc}}{\delta B_2} ~,
\end{aligned}
\label{eomsugra}
\end{equation}
where we have imposed self-duality of $F_5$ on shell.  By comparing
the equations with the Bianchi identities of the dual field strengths
we get the relation
\begin{equation}
F_7 = -~ e^\phi \ast F_3~. 
\end{equation}
Then the BI corrected by D-brane sources are
\begin{equation}
dF_3 = - 2\kappa^2 \frac{\delta S_{loc}}{\delta C_6} \qquad\qquad dH_3
= 0 ~.
\end{equation}
Remark that in our conventions, the complex 3-form $G_3= dC_2+ \tau H_3$
is simply \be G_3= F_3+i H_3 \ee when the axio-dilation is constant.

\section{Poisson equation on the singular conifold} \label{laplacian conifold}

The Poisson equation for the warp factor on the conifold reads
\begin{equation}
\bigg[ \frac{1}{r^5} \partial_r r^5 \partial_r + \frac{1}{r^2}
\sum_{i=1}^2 \Big[ \frac{6}{\sin\theta_i} \, \partial_{\theta_i}
\sin\theta_i \, \partial_{\theta_i} + 6 \Big( \frac{1}{\sin\theta_i}
\, \partial_{\phi_i} - \cot\theta_i \, \partial_\psi \Big)^2 \Big] +
\frac{9}{r^2} \, \partial_\psi^2 \bigg] h = \frac{C}{r^6} \, \delta's
\end{equation}
where the RHS is the same as in \eqref{eq warp}. Due to the symmetries
of the configuration with \Nugual{2} branes at the tip, the ansatz for
the warp factor does not depend of $\psi$ and $\phi_i$. Then we are
left with
\begin{equation}
\bigg[ \frac{1}{r^5} \partial_r r^5 \partial_r + \frac{1}{r^2}
\sum_{i=1}^2 \frac{6}{\sin\theta_i} \, \partial_{\theta_i}
\sin\theta_i \, \partial_{\theta_i} \bigg] h = \frac{C}{r^6} \,
\delta's~.
\end{equation}
Following \cite{Herzog:2004tr}, we propose an ansatz
\begin{equation}
h = \frac{1}{r^4} \, g(t, \theta_1, \theta_2) \qquad\qquad t = \log
\frac{r}{r_0}
\end{equation}
with which the Laplacian simplifies to
\begin{equation}
\Delta\, h = \frac{1}{r^6} \Big\{ -4 \partial_t g + \partial_t^2 g +
\sum_{i=1}^2 \frac{6}{\sin\theta_i} \, \partial_{\theta_i}
\sin\theta_i \, \partial_{\theta_i} g \Big\} \,.
\end{equation}
Some solutions are $g = Q+ A\, t - C f(\theta_1, \theta_2)$ and the
equation reduces to
\begin{equation}
-C \sum_{i=1}^2 \frac{6}{\sin\theta_i} \, \partial_{\theta_i}
\sin\theta_i \, \partial_{\theta_i} f = 4A + C \, \delta's~.
\end{equation}
The constant $Q$ is related to a $\delta(r)$ that is the number of
D3-branes at the tip.  In \cite{Herzog:2004tr} a constraint relation
between $A$ and $C$ is found, which amounts to charge cancellation on
the compact angular sections. We will not care about it here, and
simply try to find solutions.

It will prove useful to introduce Legendre polynomials, which are
eigenfunctions of the angular Laplacian%
\footnote{We only write the relevant part including derivatives with
respect to $\theta_i$.}
\begin{align}
&\Delta_{ang} = \sum_{i=1}^2 \frac{6}{\sin\theta_i} \,
\partial_{\theta_i} \sin\theta_i \, \partial_{\theta_i}  = 6
\sum_{i=1}^2 \partial_{\cos\theta_i} (1-\cos^2\theta_i)
\partial_{\cos\theta_i} \\ &\Delta_{ang} \,P_n(\cos\theta_i) = -6
n(n+1) \, P_n (\cos\theta_i)\; \qquad (i=1,2)~.
\end{align}
The last formula follows from the differential equation
\begin{equation}
(1-x^2)P_n''(x)-2xP_n'(x) +n(n+1)P_n(x)=0~.
\end{equation}

The eigenfunctions of the angular Laplacian on the conifold are
products of Legendre polynomials
\begin{equation}
\Delta_{ang} \, P_{l_1}(\cos\theta_1) P_{l_2}(\cos\theta_2) = -6 \big[
l_1(l_1+1) + l_2(l_2 +1) \big] \; P_{l_1}(\cos\theta_1)
P_{l_2}(\cos\theta_2) \,.
\end{equation}
The product of $\delta$-functions is easily written as
\be
4 \delta(1-\cos\theta_1) \delta(1-\cos\theta_2) = \sum_{l_1=0}^\infty (2l_1+1) \, P_{l_1}(\cos\theta_1) \,
\sum_{l_2=0}^\infty (2l_2+1) \, P_{l_2}(\cos\theta_2) ~.
\ee
Then the solution we are looking for is
\begin{equation}
f = \frac{1}{24}{\sum_{l_1, l_2 \neq (0,0)}^\infty } \frac{(2l_1
+1)(2l_2+1)}{l_1(l_1+1) + l_2(l_2+1)} \, P_{l_1}(\cos\theta_1) \,
P_{l_2}(\cos\theta_2)\;,
\end{equation}
where this last sum excludes $(l_1,l_2)=(0,0)$. One gets
\begin{equation}
\Delta_{ang} \, f = -\delta(1-\cos\theta_1) \delta(1-\cos\theta_2)
+ \frac{1}{4} ~.
\end{equation}


\section{Periods of $\Omega$}\label{compute Omega periods}

Here we provide some details on the computation of the periods of
$\Omega$ in the deformed orbifolded conifold.  A general expression
for the holomorphic 3-form is given by  
\be 
\Omega \propto \frac{1}{2\pi i} \oint_{P=0} \frac{dw_1 \wedge dw_2 \wedge dw_3 \wedge
dw_4}{P} = \frac{dw_1 \wedge dw_2 \wedge dw_3}{\partial P / \partial
w_4}~, 
\ee 
where $P[w]$ is the polynomial equation defining the
geometry. We take 
\be
\label{Pdefocon} P = xy - (u^2 - v^2 +
\epsilon_1)(u^2 - v^2 + \epsilon_3) = 0~.  
\ee 
The geometry is
described as in appendix \ref{sec: 3-cycles}: the cylinder $xy=\mathrm{const.}$ is
fibered over $\bbC^2\cong \{u,v\}$.  The fibration degenerates at the
loci 
\be
\label{u1u2blabla} u_1^2 = v^2 - \epsilon_1, \qquad
\mathrm{and} \qquad  u_2^2 = v^2 - \epsilon_3, 
\ee 
and the 2- and 3-cycles are visualised as in Fig.\ref{fig: def geometry cycles}.

Choosing a convenient normalisation, we have 
\be 
\Omega =
\frac{1}{2\pi^2} \, \frac{du\wedge dv \wedge dx}{x} ~.  
\ee 
Then, for any 3-chain $\Pi_3$ 
\be 
\int_{\Pi_3} \Omega = \frac{i}{\pi} \int_{C_j} du \wedge dv = \frac{i}{\pi} \int_{\gamma_j} u\, dv ~.  
\ee
Here $C_j$ is a 2-chain over which an $S^1$ is fibered according to
(\ref{Pdefocon}), giving us the 3-chain, and  $\gamma_j$ is its
boundary.  The geometry is then visualized as a double-sheeted
$v$-plane, with the upper and lower sheets connected through the cuts
at $u_1^2=0$ and $u_2^2=0$ (see \ref{u1u2blabla}).

Then the 3-cycle $A_i$ corresponds to $\gamma_i$ circling around the
corresponding cut on the $v$-plane, while for $B_i$ one goes from the
upper sheet to the lower one through the cut. Using the indefinite
integral  
\be 
F(v, \epsilon) \equiv \int\sqrt{v^2-\epsilon} \;dv =
\frac{1}{2}\Big[ v  \sqrt{v^2 - \epsilon} -  \epsilon \log \big( v +
\sqrt{ v^2 - \epsilon} \big) \Big] ~, 
\ee 
whose expansion for $v^2 \gg \epsilon$ goes as 
\be 
F(v, \epsilon) = \frac{1}{2}v^2 - \frac{1}{4}
\epsilon \log{ (4v^2e)}  + \orders{\frac{\epsilon^2}{v^4}} ~, 
\ee 
we obtain 
\be 
\int_{A_j} \Omega = \epsilon_j, \qquad \text{and} \qquad
\int_{B_j} \Omega  =\  \frac{\epsilon_j}{2\pi i}  \log \left(
\frac{\epsilon_j}{4ev_0^2}\right) + \text{regular} ~, 
\ee 
where $v = v_0$ is a cut-off for the non-compact cycle.

Similarly, we can consider a 3-chain $\Xi_3$ that begins on a
representative of $\calC_2$ stretching between $u_1=\xi$ and $u_2=\xi$
in $\bbC^2=\{u,v\}$, and goes to infinity at $v=v_0$.  For $|v_0|^2
\gg |\epsilon_k|$, the integral of $\Omega$ over $\Xi_3$  is (notice
that contrarily to what happens for the $B$-cycle we do not integrate
past the cut) 
\be 
\int_{\Xi_3} \Omega = \frac{1}{2\pi i} \Big[
F(\xi,\epsilon_1) - F(\xi,\epsilon_3) + (\epsilon_1 - \epsilon_3) \log
(2e^{1/4}\, v_0) \Big] + \text{regular}~.  
\ee 
In the limit $|\xi|^2 \gg |\epsilon_k|$, we get the simpler result (\ref{simplePerXi3}).


\end{document}